\documentclass[prx,twocolumn,aps,epsf,showpacs,superscriptaddress,longbibliography,nofootinbib,notitlepage]{revtex4-2}
\usepackage{graphicx}
\usepackage{latexsym,amsmath,mathtools,amssymb,bbm,array,comment}
\usepackage[normalem]{ulem}
\usepackage[dvipsnames]{xcolor}
\usepackage[colorlinks=true,allcolors=blue]{hyperref}

\usepackage{titlesec}
\titleformat{\paragraph}[runin]
        {\itshape}
       {}
        {0pt}
        {}
        [ -]
\titlespacing*{\paragraph}{0pt}{6pt}{4pt}

\newcommand{\<}{\langle}
\newcommand{\up}{\uparrow}
\newcommand{\down}{\downarrow}

\renewcommand{\>}{\rangle}
\renewcommand{\(}{\left(}
\renewcommand{\)}{\right)}
\renewcommand{\[}{\left[}
\renewcommand{\]}{\right]}

\renewcommand{\d}{\partial}

\newcommand{\Z}{\mathbb{Z}}
\newcommand{\T}{\mathcal{T}}

\newcommand{\phpf}{{\rm PHPf^\up\times \overline{PHPf}^\down}}
\newcommand{\pf}{{\rm Pf^\up\times \overline{Pf}^\down}}
\newcommand{\ueight}{{\rm U(1)_8^\up\times U(1)_{-8}^\down}}
\newcommand{\tbmo}{tb-MoTe$_2$}
\newcommand{\gsd}{{\rm GSD}}
\newcommand{\moire}{moir\'e}
\usepackage{centernot}

\newcommand{\e}{e}
\newcommand{\m}{m}

\newcolumntype{C}[1]{>{\centering\arraybackslash}m{#1}}
\AtBeginDocument{
\heavyrulewidth=.08em
\lightrulewidth=.05em
\cmidrulewidth=.03em
\belowrulesep=.65ex
\belowbottomsep=0pt
\aboverulesep=.4ex
\abovetopsep=0pt
\cmidrulesep=\doublerulesep
\cmidrulekern=.5em
\defaultaddspace=.5em
}
\usepackage{booktabs}
\newcolumntype{R}[1]{>{\raggedleft\arraybackslash}p{#1}}
\AtBeginDocument{
\heavyrulewidth=.08em
\lightrulewidth=.05em
\cmidrulewidth=.03em
\belowrulesep=.65ex
\belowbottomsep=0pt
\aboverulesep=.4ex
\abovetopsep=0pt
\cmidrulesep=\doublerulesep
\cmidrulekern=.5em
\defaultaddspace=.5em
}

\newcommand{\red}[1]{\textcolor{red}{#1}}

\definecolor{bleudefrance}{rgb}{0.19, 0.55, 0.91}

\newcommand{\TO}{TO~}

\begin{document}
\title{Cheshire qudits from fractional quantum spin Hall states in twisted MoTe$_2$}
\author{Rui Wen}
\author{Andrew C. Potter}
\affiliation{Department of Physics and Astronomy, and Stewart Blusson Quantum Matter Institute, University of British Columbia, Vancouver, BC, Canada V6T 1Z1}

\begin{abstract}
Twisted MoTe$_2$ homobilayers exhibit transport signatures consistent with a fractional quantum spin Hall (FQSH) state.
We describe a route to construct topological quantum memory elements, dubbed Cheshire qudits, formed from punching holes in such a FQSH state and using proximity-induced superconductivity to gap out the resulting helical edge states.
Cheshire qudits encode quantum information in states that differ by a fractional topological ``Cheshire" charge that is hidden from local detection within a condensate anyons.
Control of inter-edge tunneling by gates enables both supercurrent-based readout of a Cheshire qudit, and capacitive measurement of the thermal entropy associated with its topological ground-space degeneracy.
Additionally, we systematically classify different types of gapped boundaries, Cheshire qudits, and parafermionic twist defects for various Abelian and non-Abelian candidate FQSH orders that are consistent with the transport data, and describe experimental signatures to distinguish these orders.
\end{abstract}

\maketitle 

Recent experiments~\cite{kang2024evidence} in twisted homobilayer MoTe$_2$ (\tbmo) at band filling fraction $\nu=3$ show electrical transport signatures that are broadly consistent with an even-denominator (per spin) fractional quantum spin Hall (FQSH) state, including vanishing Hall conductance, and approximately quantized two-terminal edge electrical conductance $\sigma_{2T} = \frac{3}{2}\left(\frac{2e^2}{h}\right)$.
The edge transport signatures are broadly consistent with multiple candidate Abelian- and non-Abelian FQSH orders~\cite{li2021spontaneous,jian2024minimal}, that all exhibit anyons with fractional charge- and $S^z$-spin, and helical edge states protected by symmetries associated with charge- and spin/valley- conservation and time-reversal.
%
Distinguishing which of these candidate orders is realized in \tbmo remains an outstanding basic question.

\begin{figure}[t!]
\includegraphics[width = 0.5\textwidth]{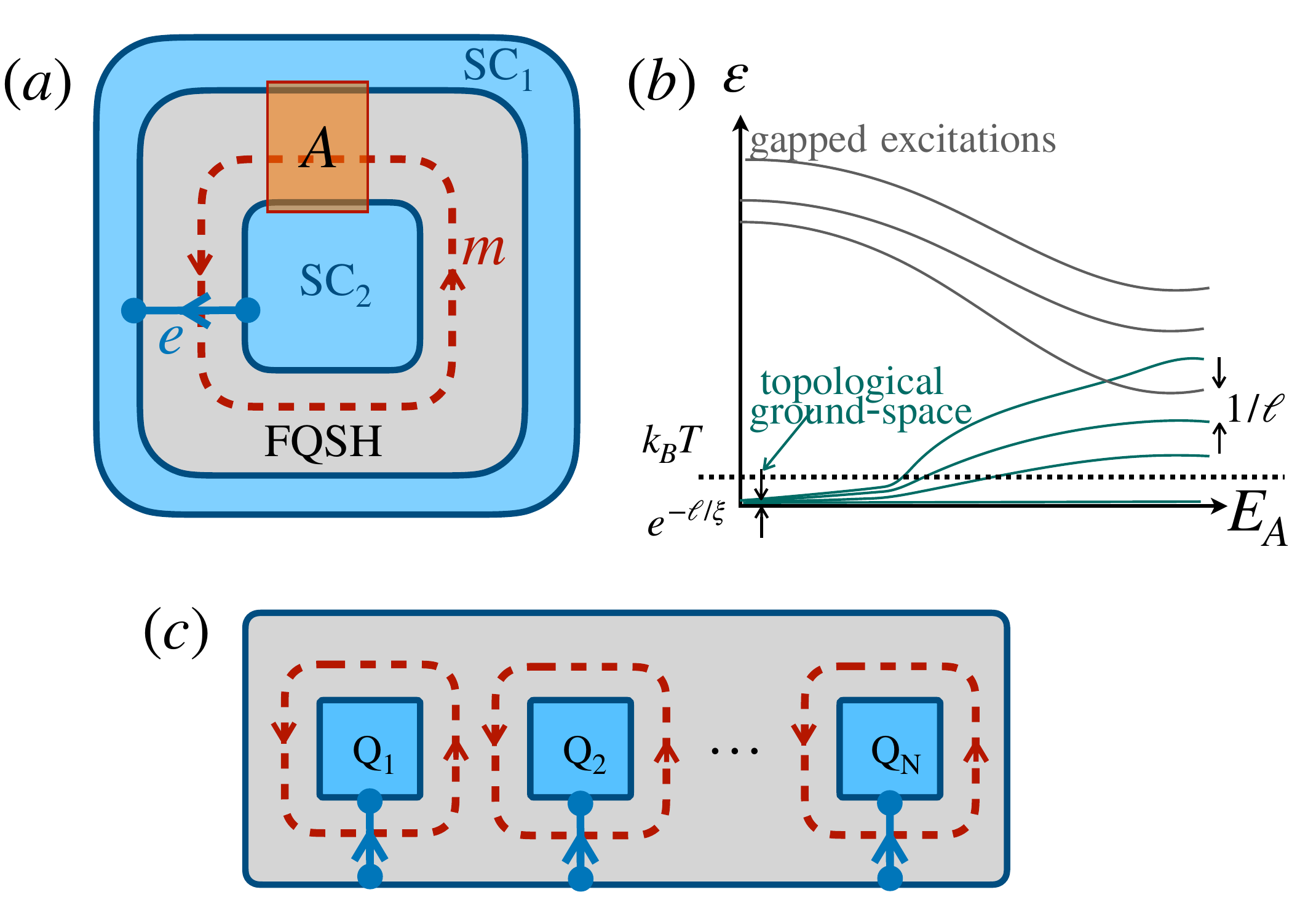}
\caption{{\bf Superconducting Cheshire qudit  -- } (a) A Cheshire qudit made from an annulus of FQSH state with boundaries gapped by proximity to conventional superconductors. The resulting system has a topological ground-state degeneracy (GSD), with different characterized by a fractional, non-locally encoded ``Cheshire" charge of the edge.
The logical operators of the qudit are implemented by tunneling of fractional-charge, $e$, anyons between the edges (blue line ending in dots), and braiding fractional spin, $m$, anyons around the inner hole (red dashed line).
Inter-edge tunneling for detection and readout are controlled by an electric field applied by dual gates (orange square) in region $A$. (b) Schematic of the energy spectrum of the Cheshire qudit as a function of the perpendicular electric field, $E_A$ in the gate region. Increasing $E_A$ turns on inter-edge tunneling of fractional charge $e$ particles, selectively lifting the GSD, and enabling supercurrent readout of the qudit state or entropic measurement of the GSD. (c) 
Schematic of array of $N$ Cheshire charge qudits, $Q_{1,\dots N}$ and associated logical operators made from $N$ superconducting islands in an FQSH sheet.
}
\label{fig:cheshire}
\end{figure}

Apart from the intrinsic scientific interest in realizing a fundamentally new fractional topological insulator phase, these observations raise the prospect of engineering topologically-protected qudits that are protected from local noise and decoherence.
Previous theoretical proposals focused on engineering non-Abelian parafermionic twist defects (generalizing Majorana bound states) from complex  triple-junction heterostructures of Abelian FQSH edge states, superconductors, and magnetic insulators~\cite{alicea2016topological}.
In this work, we propose an alternative and potentially simpler route to topological qudits based only on local electrostatic gating and proximity-induced superconductivity without magnetic elements, which adapts the ideas introduced abstractly for general Abelian fractional topological insulators~\cite{iadecola2014accessing}, and fractional charge superconductors in electron-hole bilayers~\cite{barkeshli2016charge} to a realistic experimental platform. The architecture (Fig.~\ref{fig:cheshire}(a)) involves using local gates, or depositing superconducting Islands,  
to punch holes in a sheet of FQSH order. Proximity to a conventional superconductor produces a condensate of charged anyons that gaps out the helical edge states propagating around the edge of each hole, leaving only a global topological ground-state degeneracy (GSD), which serves as a topologically-protected quantum memory~\cite{iadecola2014accessing,barkeshli2016charge}. 
The GSD of each hole is encoded in a fractional electrical charge, dubbed ``Cheshire charge"~\cite{alford1990interactions,preskill1990local,barkeshli2016charge}, which is hidden by quantum fluctuations of a condensate of charged anyons, and  cannot be locally detected.
Correspondingly, we dub the resulting device, a Cheshire qudit ($d$-level quantum memory whose dimension, $d$, depends on the underlying FQSH order).

%
%
%
Local gating can be used to control tunneling between the edges of two holes, which enables a supercurrent-based readout of Cheshire qudits~\cite{barkeshli2016charge,repellin2018numerical}. In addition, we propose a means to detect the topological ground-space degeneracy (GSD) of Cheshire qudits via an associated thermal entropy using all-electrical measurements of polarization.
This enables one to directly distinguish Abelian from non-Abelian FQSH candidate orders, without measuring thermal transport, and simplifies the experimental protocols compared to previously proposed means to entropically measure ground-state degeneracy of non-Abelian defects~\cite{cooper2009observable,iadecola2014accessing,sankar2023measuring}.

For concreteness, in the main text, we illustrate these ideas for the simplest Abelian candidate $\Z_4$ FQSH order~\cite{potter2017realizing,sodemann2017composite,jian2024minimal}, with superconducting boundaries implemented by proximity to a conventional superconductor. 
It is also possible to gap the edge of the FQSH state by breaking spin/valley conservation and time-reversal, which leads to a dual type of Cheshire spin qudit. However, due to the strong spin orbit coupling in TMD materials, this \emph{cannot} be simply implemented by an in-plane magnetic field, but rather requires a spontaneous development of inter-valley coherent (IVC) order with an in-plane spin density with characteristic wave vector $2K$, and zero net in-plane magnetic moment. 
In the supplemental materials~\cite{suppinfo}, review of other candidate FQSH orders and their edge state physics, and provide a systematic classification of their non-Abelian defects and Cheshire qudits.
We find that, for all the candidate orders, the different gapped boundaries and twist defects are in one-to-one correspondence with those of the Abelian $\Z_4$ order studied in the main text, though the Cheshire qudits dimensions and logical operator structure differ depending on the underlying FQSH phase. In particular, the twist defects are all either $\Z_4$ parafermions or Majorana bound-states, even when the underlying FQSH order is non-Abelian.

\paragraph{Symmetries and Band-Structure}
In \tbmo, the z-component of spin is energetically locked, to the valley degree of freedom by a strong spin-orbit coupling such that the first valence band has spin-up (down) electrons in the $K$ ($\text{-}K$) valley respectively.
The resulting bands have symmetries associated with charge (c) conservation, time-reversal (TR), and an approximate emergent spin/valley (sv) conservation leading to overall symmetry group $\[U(1)_c \rtimes \Z_2^T\]\times U(1)_{sv} $.
Topological band-insulators with this symmetry are classified by an integer-valued quantum spin Hall (QSH) index, that indicates the Chern number for spin-up/$K$-valley electrons (opposite to that of the spin-down/$(\text{-}K)$-valley electrons).

Electronic structure calculations for \tbmo{} at twist angles near $2^\circ$ reveal multiple, isolated, narrow-bandwidth QSH bands arising from a \moire{} skyrmion crystal pattern for the inter-layer pseudospin~\cite{wu2019topological,devakul2021magic,zhang2021spin}. 
We assume that state observed at band filling $\nu=3$  arises from a combination of a completely filled and essentially inert QSH band, and a half-filled QSH band that is driven into a fractionalized state by strong interactions.

The electrical transport signatures observed in~\cite{kang2024evidence} are consistent with multiple Abelian- and non-Abelian topological orders~\cite{jian2024minimal} for the half-filled QSH band. 
Each candidate exhibits symmetry-protected edge states~\cite{jian2024minimal,may2024theory,chou2024composite} that yield quantized electrical two-terminal edge conductance $\sigma_{2T} = 3\frac{e^2}{h}=\frac{3}{2\pi}$  (throughout, we adopt natural units with $k_B,\hbar,e=1$), but potentially differ in their thermal edge conductance (see Appendix~\ref{app: glossory}).
In the main text, we focus primarily on the simplest Abelian candidate state with $\Z_4$ topological order originally proposed in~\cite{potter2017realizing,sodemann2017composite}, and recently revisited in the context of \tbmo~\cite{jian2024minimal}. This FQSH order can be viewed as a quantum-disordered version of an inter-valley coherent (IVC) state. The IVC state is characterized by spontaneous breaking of time-reversal and $S^z$ rotation symmetries through inter-valley spin exciton condensation with order parameter: $\<c^\dagger_{\up,K}c^{\vphantom\dagger}_{\down,-K}\> = |\Delta_{IVC}|e^{i\phi}$. Microscopically, the IVC order corresponds to a wave vector $2K$ spin-density wave with colinear spin texture along a spontaneously-chosen in-plane direction at angle $\phi$. 
Flavor polarized states such as $S^z$-polarized ferromagnets or IVC states are favored by exchange energies in extremely flat and well-isolated bands.
Empirically, the FQSH state observed in tb-MoTe$_2$ must arise in a regime where the TI bands are narrow enough that interactions can drive the system into a correlated insulator, 
but where the order is sufficiently frustrated by residual band-width and band-mixing effects to evade such topologically-trivial spontaneous symmetry broken states, which do not have conducting edge states.

One route to locally gain from the exchange interactions, without breaking symmetry is for quantum fluctuations to disorder the IVC order by proliferating vortices of the IVC phase, without killing off the local IVC order amplitude (and its associated gap for electrons).
Due to the QSH response the elementary $2\pi$-vortex, $\e$, of the IVC order binds $1/2$ electric charge.
Consequently, the minimal vortex condensation consistent with symmetry and experimental observations is a $8\pi$-vortex condensate, which fractionalizes the IVC order parameter field, $\m=e^{i\phi/4}$, that is one-quarter of the IVC order parameter, i.e.  carries fractional spin $S^z=1/4$.
Since the fractionalized order parameter sees the elementary vortex as a $2\pi/4$ flux, adiabatically dragging an $\m$ around an $\e$ results in a mutual anyonic-exchange phase of $e^{2\pi i/4}$.
The resulting topological order is that of a $\Z_4$-gauge theory with $\e,\m$ being the gauge charge- and magnetic flux excitations respectively.

\paragraph{Gapped edges and Cheshire qudits}
This $\Z_4$-FQSH order exhibits helical edge states described by a Luttinger-liquid Hamiltonian~\cite{alicea2016topological}:
\begin{align}
H_{LL} = \sum_{I,J\in \{\e,\m\}} \frac{v}{4\pi} \int dx \d_x\phi_I G_{IJ} \d_x \phi_J\label{eq: HLL}
\end{align}
where $x$ is a coordinate along the edge, $e^{i\phi_{\e,\m}(x)}$ respectively create $\e,\m$ excitations at position $x$ along the edge, and satisfy commutation relations $[\phi_I(x),\d_{x'}\phi_J(x')] = 2\pi iK^{-1}_{IJ} \delta(x-x')$ with $K_{IJ} = 4\sigma^x_{IJ}$ where $\sigma^{x,y,z}$ are standard $2\times 2$ Pauli matrices. Here, $v$ is a characteristic edge velocity, and the interaction matrix $G_{IJ} = (1-g\sigma^z_{I,J})$ is controlled by a single dimensionless coupling constant, $-1\leq g\leq 1$, with $g>0$ ($g<0$) corresponding to repulsive (attractive) interactions. 
We note that, unlike for topological insulators with $S^z$-non-conserving spin-orbit coupling, time-reversal symmetric correlated back-scattering terms such as $\cos 2n\phi_{\e,\m}$, which could erode edge conductance quantization, are forbidden by charge- and $S^z$ conservation.

Breaking these protecting symmetries can open a gap in the edge~\cite{alicea2016topological,chou2024composite}.
For example, breaking $U(1)_c$ charge conservation by inducing edge superconductivity by proximity to a conventional superconductor induces the effective edge coupling $H_{\rm SC} = \int dx\[|\Delta|e^{i\theta_S} c^\dagger_\up c^\dagger_\down+h.c.\]$ where $\Delta$ is the proximity-induced pairing amplitude and $\theta_S$ is the superconducting phase. 
The low-energy bosonized form of this Hamiltonian reads:
\begin{align}
H_{\rm SC}&\approx -\lambda_S\int dx \cos (4\phi_{\e}(x)-\theta_S),
\end{align}
where $\lambda_S\sim \Delta_S/a_M$ with $a_M$ the \moire{} lattice spacing. This term has scaling dimension $\Delta_{e^4}=2\sqrt{\frac{1+g}{1-g}}$, and is perturbatively relevant for sufficiently attractive interactions when $\Delta_{e^4}<2$, i.e. $g<0$. Alternatively (and more realistically), this term can be made non-perturbatively relevant even for repulsive interactions by sufficiently large proximity-induced pairing $\Delta_S$.

When relevant, this term pins the phase of the edge mode to one of four degenerate minima: 
\begin{align}
\phi_{\e} = \frac{\theta_S}{4} +\frac{2\pi q}{4}, ~~~q=0,1,2,3.
\label{eq:minima}
\end{align}
Physically, these different minima correspond to different amounts of fractional charge $q/2$, referred to as Cheshire charge~\cite{alford1990interactions,preskill1990local}, since it cannot be locally detected.
This pinning of the $\phi_{\e}$ phase at an edge gives an expectation value to the creation operator for $\e$ excitations, $e^{i\phi_{\e}}$, and hence corresponds to ``condensing" $\e$ particles at the edge. Line defects that host anyon condensate are commonly referred to as Cheshire strings~\cite{alford1990interactions,preskill1990local,tantivasadakarn2024string}. Hence, we dub the associated topological quantum memory associated these degenerate minima of a superconducting edge a \emph{Cheshire qudit}.

The Hilbert space spanned by these different fractional charge states is non-locally encoded, and forms a topologically-protected quantum memory that is protected from local noise and decoherence. Namely, detecting the fractional charge for an extended superconducting edge requires measuring the phase obtained by acting with a loop operator $W^{\m}_i$ that drags an $m$ particle counterclockwise around hole, $i$, and changing it requires acting with a line-segment operator $W^{\e}_{i}$ that transfers an $e$ particle from hole $i$ to the boundary (or to another edge).
Since $e$ and $m$ are both gapped excitations, these processes are heavily suppressed for low temperature and well-separated edges by factors of $e^{-\ell/\xi}$ or $e^{-\Delta/T}$ where $\ell$ is either the circumference of the hole or distance between holes, $\Delta$ is the bulk FQSH gap and $\xi \sim \Delta/v$ is the associated correlation length, and $T$ is the temperature. 

For a disk geometry with a single superconducting edge, the constraint that the total charge and spin correspond to that for an integer number of electrons fixes a unique ground-state.
However, if there are multiple boundaries, for example by ``punching" $N$ well-separated superconducting holes into the FQSH state results in $4^N$ different possible combinations of values of the fractional electrical charge for the edge of each hole, i.e. a $4^N$-fold topological ground-space degeneracy (GSD).

\paragraph{Other Cheshire qudits and FQSH orders}
There are generally multiple distinct possible types of Cheshire qudits, corresponding to different patterns of anyon condensations that correspond to gapped boundaries. 
For example, in the $\Z_4$ FQSH state, condensing $\m$ excitations (which breaks $S^z$ and TR symmetries) results in Cheshire spin qudits that store quantum information in four different fractional spin states.
Further, interfaces between two different types of gapped boundaries along an edge host localized non-Abelian defects that are either Majorana like for $\e |( \e^2,\m^2)$ or $\m | (\e^2,\m^2)$ boundaries, or $\Z_4$ parafermions for $\e|\m$ boundary interfaces.

Apart from the $\Z_4$ topological order there are a number of other Abelian and non-Abelian candidate FQSH orders~\cite{jian2024minimal} consistent with the transport data~\cite{kang2024evidence}.
The other candidates are all \emph{helical} topological orders, consisting of a product of a \emph{chiral} topological order, $\TO$, for spin-up electrons and its time-reversed conjugate order, $\overline{\TO}$ for spin-down electrons.
In~\cite{suppinfo}, we classify all of the different possible gapped boundaries and resulting Cheshire qudits, and twist defects for each of these helical orders that have been proposed for the $\nu=3$ FQSH state in \tbmo.
We point out that inducing a symmetry-breaking gap in the edge states of a helical FQSH order on an annulus (or multi-holed sheet) is equivalent to realizing the chiral $\TO$ on a torus (or higher-genus closed surface).
We prove that, for all cases, the classification of gapped boundaries and interface (``twist") defects reduces to that of the $\Z_4$ FQSH order, but the resulting Cheshire qudits differ in their dimension and native topological gate operations depending on the underlying FQSH order.


\paragraph{Gate controlled inter-edge tunneling}
Local gating in a region $A$ that connects the inner- and outer- edges of a Cheshire qudit enables electrostatic control of the inter-edge tunneling. This capability facilitates electrical detection of the topological GSD and readout of the state of a Cheshire qudit.
Gates can control either the local control of chemical potential or electrical field. For concreteness, we focus on the latter below.

 Consider an annular geometry with inner- and outer- edges labeled by indices $1,2$ respectively (Fig.~\ref{fig:cheshire}a inset), with dual gating in region $A$.
Assuming the lowest energy bulk anyon excitation is the minimal-charge $\e$ particle, and labeling the fields of the inner- and outer- edges by indices $1,2$ respectively, then inter-edge tunneling in the gated region, $A$, is described by the effective Hamiltonian:
\begin{align}
H_{\Gamma} = -\Gamma(E_A)\int_{-\ell_A/2}^{\ell_A/2} dx \cos\(\phi_{\e,1}-\phi_{\e,2}-\frac{2\pi \Phi_A }{8\Phi_0}\frac{x}{\ell_A}\).
\label{eq:tunneling}
\end{align}
Here, $E_A$ is the out-of-plane electric field in the gate region $A$ with length $\ell_A$. For generality, we have also included the effects of an out-of-plane magnetic field, where $\Phi_A$ is the magnetic flux through region $A$, and $\Phi_0 = hc/2e=\pi$  is the superconducting flux quantum.
The tunneling amplitude is approximately described by $\Gamma(E_A) \approx \Gamma_0 e^{-\ell/\xi(E_A)}$ where $\Gamma_0$ is a tunneling amplitude, $\ell$ is the inter-edge distance, and $\xi(E_A)$ is the correlation length associated with the bulk FQSH gap. Empirically~\cite{kang2024evidence}, an electric field can weaken the FQSH gap, thereby increasing $\xi$, and eventually driving region $A$ into a topologically trivial metallic quantum dot (with finite size gap inversely proportional to its area). Hence, if $\ell/\xi(E=0) \gg 1$, the tunneling amplitude in region $A$ can be tuned over a wide range by the electric field.

\paragraph{Supercurrent readout of Cheshire qudits}
The state of the Cheshire qudit can be detected by its influence on the current-phase relation in the junction formed in region $A$~\cite{barkeshli2016charge}.
The phase difference between the inner- and outer- superconductors in region $A$ can be controlled by a perpendicular magnetic field, which sets the phase difference between the inner and outer superconductors to $\theta = 2\pi (\Phi-\Phi_A)/8\Phi_0$ where $\Phi$ is the total flux through the FQSH region. When the area of region $A$ is substantially smaller than the entire FQSH region, $\theta$ and $\Phi_A$ can be tuned approximately independently.
In the weak-tunneling regime, $\Gamma\ll \lambda_S$, the FQSH edge fields are strongly pinned to the local minima described by (\ref{eq:minima}): $(\phi_{\e,1}-\phi_{\e,2}){\rm ~mod~}2\pi \approx \frac\theta4+\frac{\pi q}{2}$.
This results in super-current, $I_S = -\frac{2\pi}{\Phi_0}\frac{\d \<H_{\Gamma}\>}{\d\theta}$:
\begin{align}
I_S \approx \frac{\pi \Gamma \ell_A}{\Phi_0} \sin\(\frac{\theta}{2}+\frac{\pi q}{2}\){\rm sinc}\( 2\pi \Phi/8\Phi_0\) 
\end{align}
with ${\rm sinc}(x) = \sin x/x$.
There are two notable features. First, the state of the Cheshire qudit can be read out by the $\pi q/2$ phase offset. Second, the flux periodicity is octupled compared to a conventional Josephson junction since the Josephson coupling is mediated by fractional charge-half particles rather than charge-two Cooper pairs.
We note that, any qudit readout scheme must be implemented faster than the effective $T_1$ relaxation time of the qudit (here due to quasiparticle poisoning by fractional charge anyons, but \emph{not} by electrons), which may require high-frequency readout schemes~\cite{rokhinson2012fractional}. 

\begin{figure}[t!]
\includegraphics[width = 0.5\textwidth]{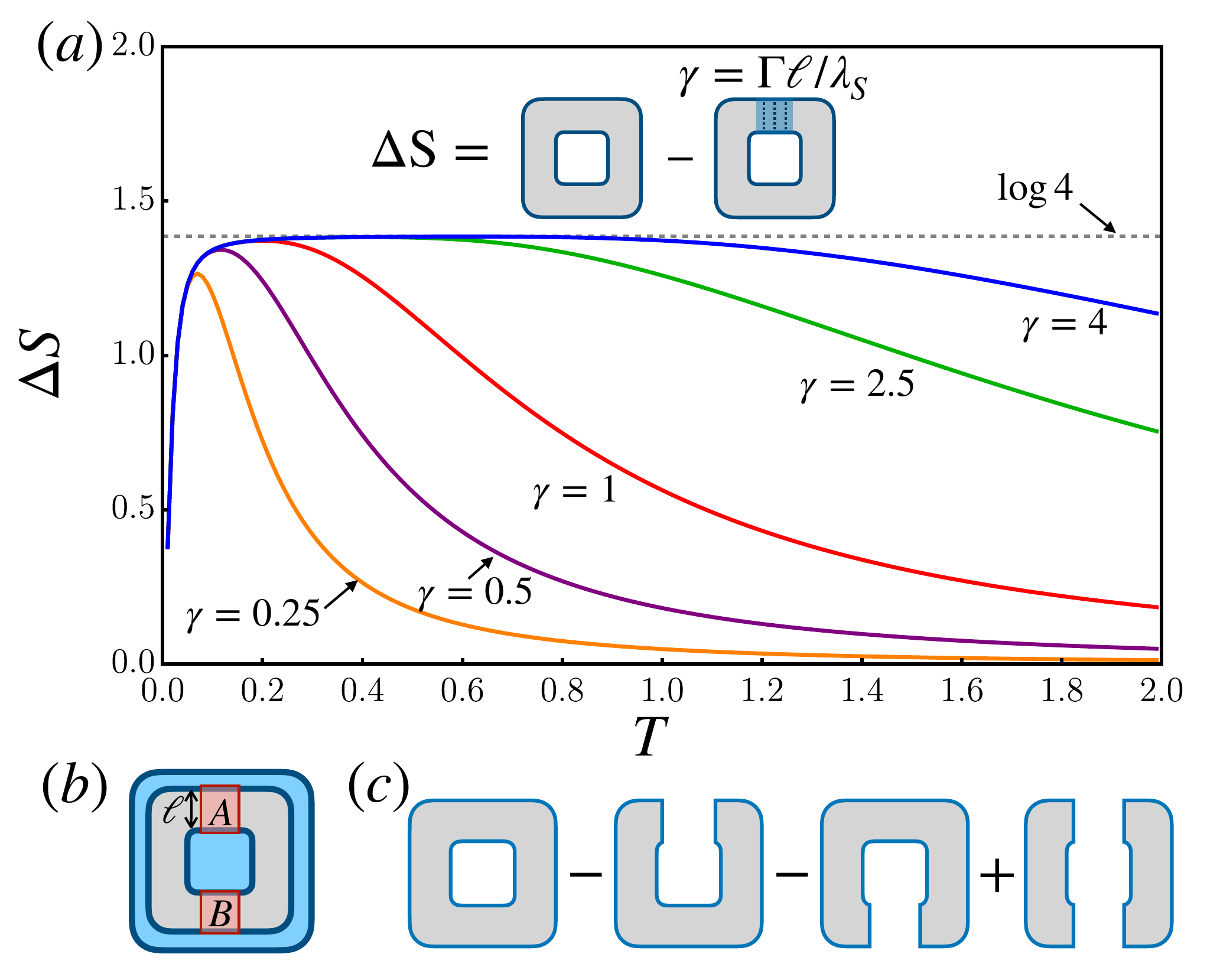}
\caption{{\bf Entropic detection of topological ground-state degeneracy  -- } 
Numerical simulation of entropy-based GSD detection scheme in a lattice discretization of the proximitized FQSH edge states.
Inter-edge tunneling strength is characterized by a dimensionless parameter $\gamma=\Gamma \ell/\lambda_S$, where $\Gamma$ is the tunneling amplitude, $\lambda_S=2$ is the proximity-induced superconducting gap, and $\ell=10$ sites is the length of the tunneling region.
Upon increasing $\gamma$ from an initial value of $0.025$ to the final values listed on the plot, $\Delta S$ exhibits a nearly-quantized plateau at value $\log {\rm GSD}$ over an appropriate temperature range described in the text.
$\Delta S$ can be detected through electrical polarization measurements exploiting a Maxwell relation. 
(b,c) As a refinement to this scheme, spurious non-topological contributions to $\Delta S$ from low-energy local bound states can be eliminated by additionally gating a distant region $B$ (b), and computing the linear combination of entropy measurements shown in (c), where gaps in the annulus indicate strong tunneling in regions $A$ and/or $B$.
}
\label{fig:entropy}
\end{figure}
\paragraph{Entropic measurement of GSD}

A Cheshire qudit in thermal equilibrium at intermediate temperatures: $\delta\ll T\ll \Delta$, that greatly exceed the (exponentially-small) splitting, $\Delta$, of the GSD, but are much less than the (bulk or edge) gap to local excitations, $\Delta,$, will be in an approximately equal-weight (incoherent) mixture of the topological ground-states, and will exhibit an associated thermal entropy $S\approx \log {\rm GSD}$, with corrections that are exponentially suppressed in $e^{-\Delta/T},e^{-\ell/\xi}$ where $\ell$ is the distance between different edges of the annulus.
Previous theoretical works proposed detecting related thermal entropies of non-Abelian defects either through calorimetry or by exploiting Maxwell relations relating changes in entropy to changes in charge or magnetization~\cite{cooper2009observable,iadecola2014accessing,sankar2023measuring}, the latter of which have been used to experimental measure entropy in quantum dot systems~\cite{hartman2018direct,child2022entropy,child2022robust}.
Here, we propose a scheme to measure the thermal entropy associated with the topological GSD of an annular Cheshire qudit through capacitive measurements of electrical polarization (Fig.~\ref{fig:entropy}), which introduces several potential experimental simplifications.



Consider the setup shown in Fig.~\ref{fig:cheshire}a.
As discussed above, applying an electric field, $E_A$, in region $A$ that connects the inner- and outer- edges can be used to switch on inter-edge tunneling, splitting the GSD  (Fig.~\ref{fig:cheshire}b). If the tunneling for the largest applied electric field, $E_{\max}$, satisfies: $\Gamma(E_{\max})\gg T$, then the electric field in region $A$ can be used to effectively switch the topology of system from that of an annulus to a disk.

This change in entropy can be observed by capacitive measurements of the out-of-plane electric polarization in region $A$, $\mathcal{P}_A$, via the Maxwell relation $\frac{\d S}{\d E_A} = \frac{\d \mathcal{P}_A}{\d T}$:
\begin{align}
\Delta S = \int_0^{E_{\rm max}} dE_A \frac{\d \<P_A\>}{\d T} \approx \log {\rm GSD} +\mathcal{O}\(e^{-\Delta/T},e^{-\ell/\xi}\).
\label{eq:DeltaS}
\end{align}
Such capacitive measurements have previously been used to detect small changes in polarization associated with a spontaneous layer polarization in graphene bilayers~\cite{young2011capacitance,hunt2017direct}, utilizing a single pair of top- and bottom- gates both to detect $P_A$ and apply the displacement field, avoiding the need to fabricate and calibrate additional charge-sensing devices~\cite{hartman2018direct,child2022entropy,child2022robust} or phase-coherent interferometric setups~\cite{sankar2023measuring} required to measure entropy via the charge(N)/chemical-potential($\mu$) Maxwell relation: $\frac{\d N_A}{\d T}=\frac{\d S}{\d \mu_A}$.
This setup also evades the difficulties of isolating the sub-extensive contributions of a 2d material compared to its 3d substrate that face calorimetric measurements of entropy ~\cite{iadecola2014accessing}.
Finally, since all edge states are fully gapped in our setup, there are no additional non-topological contributions to $\Delta S$ from thermally populated gapless edge modes, as for related schemes to entropically detect the total quantum dimension of fractional quantum Hall states~\cite{sankar2023measuring}.

We numerically model this scheme by simulating a lattice discretization of the Luttinger liquid edge of the $\Z_4$-FQSH order, by making Harmonic approximations ($-\cos 4\phi_{\e} \approx \frac{4^2}{2}(\phi_\e-\pi q/2)^2+{\rm const.}$) to superconducting- and tunneling- cosine terms in (\ref{eq:minima},\ref{eq:tunneling}, see Appendix~\ref{app:numerics} for details). The resulting change in entropy $\Delta S$ from (\ref{eq:DeltaS}), shown in Fig.~\ref{fig:entropy}, exhibits a nearly quantized plateau in the appropriate temperature range described above. In these simulations, the simulated system has the geometry of a quasi-2d strip with periodic boundary conditions along $x$, discretized into a $L=100$ site lattice. The superconducting gap strength is set to $\lambda_S=2$. We model the effect of applying a displacement electric field to region $1\le x\le 10$, as changing inter-edge tunneling strength $\Gamma$ from $\Gamma_i=0.005$ to $\Gamma_f$ ranging from 0.05 to 0.8.
$\Delta S$ exhibits a nearly-quantized plateau at intermediate temperatures that are much larger than the finite-size splitting of the GSD at $\Gamma_i$, and much smaller than the superconducting gap, and GSD-splitting induced by $\Gamma_f$.

In a fully-gapped system, the protocol above is sufficient to accurately extract $\log {\rm GSD}$.
However, in many situations additional low-energy localized bound states may arise due to disorder, vortices, or other mechanisms, and could contribute spurious non-topological thermal entropy contributions.
These can be eliminated by further gating a distant region $B$ (Fig.~\ref{fig:entropy}b), and computing the linear combination of entropies shown in Fig.~\ref{fig:entropy}c, which is designed to cancel local contributions, leaving only those from the global topological GSD.
This combination is very similar to constructions for extracting topological entanglement entropy (TEE)~\cite{levin2006detecting}, though we emphasize that TEE and Cheshire-qudit GSD are generally distinct quantities.


Importantly, precisely the same setup enables entropic measurement of the GSD of \emph{any} FQSH order, and provides a means to directly distinguish different Abelian and non-Abelian candidates for the $\nu=3$ sate in \tbmo. However, as with thermal conductance measurements it remains incapable of distinguishing the subtle differences between the candidate non-Abelian FQSH orders, which all have the same GSD=6.

\paragraph{Discussion}
The above proposals outline a path to creating, detecting, and reading out a Cheshire-qudit topological memory made from FQSH states interfaced with superconductors.
To implement this scheme, several experimental challenges remain. Firstly, the linear dependence of the Hall resistance on filling $\nu$ observed near $\nu=3$ in~\cite{kang2024evidence} possibly suggests that the FQSH gap in tbMoTe$_2$ is small, or that mobile bulk carriers coexist with the observed edge conduction channels.
While a small FQSH gap would be harmful to the Cheshire qudit robustness, additional non-fractionalized bulk charge carriers need not harm the qudit operation if they are gapped out by the superconducting proximity effect.
We speculate that these non-idealities might be mitigated by improved sample quality, going to lower base temperatures, or careful twist engineering to optimize the FQSH gap, however further experimental investigations are needed.

Our scheme for electronic measurement of GSD provides a potentially valuable probe of the underlying FQSH order, independent of any potential quantum information processing applications. Moreover, this technique can be more broadly applied to a wide class of 2d gate-tunable materials, to electronically detect topological order. This could provide a valuable signature, particularly for gapped quantum spin liquids such as those proposed in~\cite{law20171t, zhang2020quantum, zhang20214}, which otherwise appear featureless in electrical transport measurements.

Several challenges remain to promote this quantum memory to a fault-tolerant quantum computing architecture. 
Topological gates could be implemented by adiabatic transport of anyonic excitations around- or between- the holes defining Cheshire qudits, or possibly through measurement-only based schemes. 
It would also be interesting to look for parameter regimes (e.g in twist angle, filling fraction, or proximity to other materials) where superconductivity or IVC order could be induced within the same \moire{} heterostructure as the FQSH state. This would enable gate-based manipulation of the Cheshire qudit holes, possibly expanding the range of available topological quantum operations.
For any of the candidate FQSH orders, the resulting set of topological gates would be insufficient for universal quantum computing, and would require additional code switching or magic state distillation.
However, as for other platforms such as Majorana bound-state qubits, the intrinsic topological protection of the Cheshire qudits could still provide an advantage by reducing the overheads required to achieve a fault-tolerant threshold.

\vspace{4pt}\noindent{\it Acknowledgements} We thank Marcel Franz, Nitin Kaushal, Alberto Nocera, and Mike Zaletel for insightful discussions. This research was supported by the Natural Sciences and Engineering Research Council of Canada
(NSERC), and was supported in part by grant NSF PHY-2309135 to the Kavli Institute for Theoretical Physics (KITP).

\bibliography{PRL_v2}

\appendix

\section{Details of numerical simulation of the entropy measurement scheme\label{app:numerics}}

Here we present details behind the numerical results reported in the main text regarding the entropy measurement scheme. We model the edges of the annulus geometry Fig~\ref{fig:cheshire}a by two copies of Luttinger liquids, one for each edge. Adding proximity-induced pairing term and inter-edge tunneling term, we have the following Hamiltonian for the edge. 
\begin{align}
    H_{\text{edge}}&= \frac{v}{4\pi} \sum_{I,J\in \{\e,\m\},s} \int dx \d_x\phi_I^s G_{IJ} \d_x \phi_J^s- \nonumber\\
    &-\int_0^Ldx\sum_{s=1,2}\lambda_S \cos 4\phi_e^s-\nonumber\\
    &-\Gamma \int_{l_1}^{l_2} \cos(\phi_e^1-\phi_e^2)\label{appeq:tunnelingH}
\end{align}
$s=1,2$ labels the two edges. The length of the boundary is taken to be $L$, $\lambda_S$ is set by the amplitude of the proximity-induced pairing, and the inter-edge tunneling is assumed to take place in region $[0,\ell_A]$, with strength set by the applied perpendicular electric field. 

In the absence of  the inter-edge tunneling term, the ground state of the above model is set by pining the fields $\phi_I$ to minimize the pairing term $\lambda_S\cos 4\phi_e^s$, which leads to the ground-states: $\phi_e^1=\frac{i\pi n_1}{2},\phi_e^2=\frac{i\pi n_2}{2},~n_1,n_2=0,1,2,3$. There is a global constraint that the total charge of the annulus is an integer, which means only the four states: $|n\>:=\sum_m |\phi_e^1=\frac{\pi(n+m)}{2},\phi_2^2=\frac{\pi m}{2}\>$ are physical. Denote the perturbation of the $\phi_e^s$ fields from the $|n\>$ vacuum by $\delta\phi^s$. To quadratic order in $\delta\phi^s$, the Hamiltonian~\eqref{appeq:tunnelingH} reduces to
\begin{align}
    H_\text{edge}\approx& \int_0^L dx\sum_{s=0,1}\left( \frac{v}{4\pi} \partial_x\delta\phi_I^s G_{IJ} \d_x \delta\phi_J^s+8\lambda_S (\delta\phi^s)^2 \right)\nonumber \\
    &+\Gamma \int_{0}^{\ell} dx \left(\sin\left(\frac{\pi n}{2}\right)\delta\phi_-+\frac{1}{2}\cos\left(\frac{\pi n}{2}\right)(\delta\phi_-)^2\right)\nonumber\\
    &-\Gamma\ell_A\cos\frac{\pi n}{2}\label{eq:reducedH}
\end{align}
where we defined $\delta\phi_-:=\delta\phi^1-\delta\phi^2$. The Hamiltonian \eqref{eq:reducedH} can now be viewed as a system of finite number of bosonic particles coupled by quadratic potentials by discretizing the coordinate $x$. We numerically find harmonic modes of the system~\eqref{eq:reducedH} and compute the thermal entropy.

Defining $\delta\phi_\pm:=\delta\phi^1\pm \delta\phi^2$, the fields $\delta\phi_\pm$ decouple, and the Hamiltonian of field $\delta\phi_+$ does not depend on $\Gamma$. Thus for the purpose of extracting entropy differences for different values of applied electric field, it suffices to consider only the $\delta\phi_-$ field. The potential part of the Hamiltonian of the $\delta\phi_-$ field is 
\begin{align}
  V(\delta\phi_-)&=  \int_0^L dx \left(-\delta\phi_-\partial_x^2 \delta\phi_-+8\lambda_S (\delta\phi_-)^2\right)\nonumber\\
    &+\Gamma\int_{0}^{l_A}  \left[\sin \left(\frac{\pi n}{2}\right) \delta\phi_-+\frac{1}{2}\cos\left(\frac{\pi n}{2}\right)(\delta\phi_-)^2\right]\nonumber\\
    &-\Gamma \ell_A\cos\frac
    {n\pi}{2}.\label{eq:phipotential}
\end{align}
By discretizing $[0,L]$ into $N$ points , we may write the field $\delta\phi_-$ as a vector $\vec{\delta\phi_-}:=\{\delta\phi_-(x_i)|i=1,2,\cdots, N\}$. The Laplacian $\partial_x^2$ is then replaced by discrete Laplacian 
\begin{align}
    &(\partial_x^2 \delta\phi_-)(x_i):=\delta\phi_-(x_{i+1})+\delta\phi_-(x_{i-1})-2\delta\phi_-(x_i).
\end{align}
The potential~\eqref{eq:phipotential} can then be written as a quadratic function in $\vec{\delta\phi_-}$,
\begin{align}
   & V(\delta\phi_-)=\vec{\delta\phi_-}\cdot M\cdot \vec{\delta\phi_-}+\vec{L}\cdot \vec{\delta\phi_-}-\Gamma \ell_A\cos\frac{\pi n}{2},\nonumber\\
    &M_{ij}=-(\partial_x^2)_{ij}+8\lambda_S\delta_{ij}+\frac{\Gamma}{2}\cos\frac{\pi n}{2} \delta_{ij}\delta_{1\le i\le \ell_A},\nonumber\\
    &L_i=\Gamma \sin\frac{\pi n}{2}\delta_{1\le i\le \ell_A}.
\end{align}
We then make a change of variable, 
\begin{align}
    \vec{\delta\phi_-}\to \vec{\delta\phi_-}-\frac{1}{2}M^{-1}\cdot \vec{L},
\end{align}
 resulting in the potential function
\begin{align}
    V(\delta\phi_-)&=\vec{\delta\phi_-}\cdot M\cdot \vec{\delta\phi_-}-\frac{1}{4}\vec{L}\cdot M^{-1}\cdot \vec{L}\nonumber\\
    &-\Gamma \ell_A\cos\frac{\pi n}{2}.
\end{align}
We then diagonalize the matrix $M$, finding eigenvalues $\omega_n^i,~i=1,2,\cdots N$. The partition function of the system can then be calculated with the standard formula of entropy of harmonic oscillators
\begin{align}
    \mathcal{Z}_n=&\exp\left\{\frac{\Gamma}{T}\ell_A\cos\frac{\pi n}{2}+\frac{1}{4T}\vec{L}\cdot M\cdot \vec{L}\right\}\nonumber\\
    &\times\prod_i (1-e^{-\frac{\omega_n^i}{T}})^{-1}
\end{align}
Taking in account the fluctuation around all the vacuua $|n\>$, the total partition function of the edge is 
\begin{align}
    \mathcal{Z}=\sum_{n=0}^3 \mathcal{Z}_n
\end{align}
Entropy is then extracted using the relation 
\begin{align}
    S(T)=-\frac{\d F}{\d T} = \frac{\d}{\d T}(T\log \mathcal{Z}).
\end{align}
We compute $\mathcal{Z}(T)$ for $T$ ranging from 0 to 1.5, with increment $\delta T=0.01$, and then numerically estimate its $T$-derivative by finite difference method: $\partial_Tf(T_i)\approx f(T_{i+1})-f(T_i)/(\delta T)$.

We set $v=4\pi$ and $g=0$ throughout. The superconducting pairing amplitude is set by $\lambda_S=2$. The interval $[0,L]$ is discretized into $N=100$ points with lattice spacing $a=L/N$. 
The displacement electric field
is applied to region of size $\ell_A=10$. The initial inter-edge tunneling strength is $\Gamma_i=0.005$ and the final value $\Gamma_f$ varies from $0.05$ to $0.8$. 

\section{Glossary of candidate topological orders\label{app: glossory}}

\begin{table*}[t!]
        \caption{{\bf Selected properties of candidate FQSH orders: } including maximum quantum dimension $d_{max}$, minimum electric charge $Q_{\rm min}$ and spin/valley-charge $S^z_{\rm min}$ (in these units the electron has $q=-1$,$s=\frac 12$), two terminal electrical ($\sigma_{2T}$) and thermal ($\kappa_{2T}$) conductance (as multiples of those of an electronic integer quantum Hall state), and the dimension (GSD) of a Cheshire qudit.
         }
         \renewcommand{\arraystretch}{1.3}
        \begin{tabular}{rccccccl}
        \toprule
         FQSH order & $d_{\rm max}$ & $Q_{min}$ & $S^z_{min}$ & $\sigma_{2T}$ & $\kappa_{2T}$ & $D_{\rm Cheshire}$ & Remarks \\ 
        \midrule 
        $\Z_4$ & 1 & $\frac12$ & $\frac14$ & $1$ & $2$ & 4 & spin-valley entangled, Abelian\\
        %
        $\ueight$ & 1 & $\frac14$ & $\frac18$ & $1$ & $2$ & 8 & helical, Abelian \\
        $\pf$ & $\sqrt{2}$ & $\frac14$ & $\frac18$ & $1$ & $3$ & 6 & helical, non-Abelian \\
        $\phpf$ & $\sqrt{2}$ & $\frac14$ & $\frac18$ & $1$ & $3$ & 6 & helical, non-Abelian \\
        \bottomrule
        \end{tabular}
        \label{tab:orders}
\end{table*}

\subsection{Generalities}
The candidate FQSH orders are symmetry enriched topological orders (SETs) with $U(1)_c\times U(1)_{sv}\rtimes \Z_2^T$ symmetry.
The topological data of each theory can be specified by a list of topological ``charges" (superselection sectors or anyon types), $\{a\}$, fusion and braiding rules~\cite{kitaev2006anyons,bernevig2017topological,barkeshli2019symmetry}, as well as electrical- and spin-valley- charge assignments that are consistent that are consistent with the topological properties, and an action of time reversal that assigns a time-reversed conjugate $\mathcal{T}:a\mapsto \bar{a}$ to each anyon $a$. For anyons superselection sectors that are their own time-reverse conjugates, $\bar{a}=a$, one may also assign a value of $\T^2=\pm 1$ indicating whether the particle is a Kramers singlet/doublet. Time-reversal conjugates the topological spin $\theta_{\T(a)} = \theta_a^*$ of an anyon, inverts its spin/valley-charge, and preserves its electrical charge. For a detailed review, we refer the reader to~\cite{barkeshli2019symmetry}.

Via a standard Laughlin-type flux-insertion argument along, the quantized spin-Hall conductance requires the existence of a charge-neutral magnetic flux anyon with spin-half, and which has mutual statistics $e^{2\pi iq}$ with all fractionally charge $q$ particle. Similarly it also requires a charge-one spinless anyons that has mutual statistics $e^{4\pi i s}$ with all fractional spin-$s$ anyons. 

We can divide the candidate topological orders into two categories: i) helical orders that consist of a product of a chiral topological order for spin-up electrons and a time-reversed conjugate copy of this topological order for spin-down electrons, and ii) spin-valley-entangled orders that cannot be decomposed in a helical fashion. 

Apart from the spin-valley-entangled $\Z_4$ order described above and in the main text, the remaining FQSH candidates are helical orders, that factorize into product of a \TO for up electrons and its time-reversed \TO of down electrons.

\subsection{Spontaneous symmetry breaking states}
In a completely-flat (dispersionless) and half-filled QSH band, exchange interactions likely flavor polarization. Natural candidate states are spontaneous spin/valley polarized states with a net $S^z$ magnetization, or an intervalley coherent (IVC) state obtained by condensing inter-valley excitons.

The mean-field IVC state can be created as a filled-band of spin-valley polarized electrons 
$$|\Psi_{\rm IVC}(\phi)\> = \prod_{k} (c^\dagger_{n,k,\up}+e^{i\phi}c^\dagger_{n,k,\down})|\emptyset\>,$$
where the band-index $n$ is fixed to the (nearly-flat) half-filled topological insulator band.
This IVC state breaks TRS and $S^z$ conservation, and microscopically, corresponds to a wavevector $2K$ spin-density wave with an in-plane spin-texture. It preserves the modified time-reversal $\tilde{\mathcal{T}} = e^{-i\pi S^z}\mathcal{T}$, which is non-Kramers ($\tilde{\mathcal{T}}^2=1$).

Both spontaneous symmetry broken states are incompatible with the observations in~\cite{kang2024evidence}. Namely, the spin polarized state would exhibit a unit quantized anomalous Hall conductance, and the IVC state is topologically trivial (there are no non-trivial topological insulators with only $U(1)_c\rtimes \tilde{\mathcal{T}}$ symmetry) and does not exhibit quantized edge conductance.
Therefore, empirically, the FQSH state observed in tb-MoTe$_2$ must arise in a regime where the TI bands are narrow enough that interactions can drive the system into a correlated insulator, but not so narrow that the system simply spontaneously breaks symmetry to either spin-polarize to form an integer anomalous Hall insulator (with non-vanishing Hall conductance in contrast to observations), or to form an IVC (which does not have topological edge states, and cannot explain the observed quantized edge conductance).

\subsection{$\Z_4$ Abelian FQSH state}
One route to locally gain from the exchange interactions, without breaking symmetry is for quantum fluctuations to disorder the IVC order by proliferating vortices of the IVC phase, without killing off the local IVC order amplitude (and its associated gap for electrons).
This state was first proposed in~\cite{potter2017realizing} (see also~\cite{sodemann2017composite} for a closely related state in a half-filled quantum Hall bilayer) and later revisited in the context of \tbmo~ by~\cite{jian2024minimal}.  
The $\Z_4$ FQSH order can be understood as descending from quantum-disordering an inter-valley coherent (IVC) state that spontaneously breaks the spin/valley-conservation and time-reversal, but preserves a combination of these symmetries.

\subsubsection{Physical interpretation}
The $\Z_4$ topological ordered state arises from proliferating topological defects (vortices) in this IVC phase to restore the symmetries
Single ($2\pi)$ vortices in the direction of $\phi$ bind charge $\pm e/2$ by the spin-Hall response, denotes these $v_\pm$. 
To see this, note that the electrons the transformed quasiparticles: $f =  (e^{i\phi/2}c^\dagger_{\up}+e^{-i\phi/2}c^\dagger_{\down})$. These carry charge but their spin is neutralized (this can be thought of as adiabatically eliminating the fluctuations in the $\phi$ phase by locally rotating electron spin). A quasiparticles obtains a $(-1)$ Berry phase when encircling the $\phi$-vortex.
Since $v_\pm$ differ by binding a fermionic quasi-particle, they have mutual semionic statistics.

Double ($4\pi)$ vortices include $v_\pm^2$ which have charge $\pm 1$, and $v_+v_-$ which is a charge-neutral fermion (due to the mutual statistics of $v_+$ and $v_-$). Condensing the charge vortices would result in a superconductor with broken $U(1)_c$ symmetry, which is not observed, and the fermion cannot be directly condensed (routes to transmuting its fermionic statistics, e.g. by flux-attachment by putting it into a quantum Hall state, require breaking time-reversal symmetry, and would not result in a state with protected edge states).

By contrast, the neutral quadruple ($8\pi$) vortex, $v_+^2v_-^2$, is a boson and can be condensed to quantum disorder the IVC exciton condensate. This higher-vortex condensation deconfines a dual boson, $b=e^{i\phi/4}$ that is a quarter of the IVC order parameter, carrying $S^z=1/4$, with mutual statistics $\theta_{b,v_{\pm}} = e^{2\pi i/4}$.

The resulting state can be identified with a $\Z_4$ topological order labeling the $\Z_4$ gauge charge as $\e=b$, and gauge flux as $\m=v_+$, which have charge and $S^z$ quantum numbers, $(Q_e,S^z_e)=(\frac 12,0)$ and $(Q_m,S^z_m)=(0,\frac14 )$.

\subsubsection{Topological order data}
\begin{table}[tb!]
        \caption{{\bf $\Z_4$ topological order: } Anyon type, quantum dimension (d), self-statistics, $\theta_a$, and symmetry properties: charge, $S^z$, and time-reversed partner $\mathcal{T}(a)$. The labels range over $\Z_4$: $j,k\in \{0,1,2,3\}$.  Total quantum dimension: $D=4$. Torus $\gsd=16$. \\
         }
        \renewcommand{\arraystretch}{1.3}
        \begin{tabular}{C{0.6in}C{0.2in}C{0.4in}C{0.6in}C{0.6in}C{0.6in}}
        \toprule
         Anyon ($a$) & $d$ & $\theta_a$ & charge ($e$) & $S^z$ ($\hbar$) & $\mathcal{T}(a)$ \\ 
        \midrule
        $e^jm^k$ & 1 & $e^{i\pi jk/2}$ & $j/2$ & $k/4$ & $e^jm^{-k}$  \\
        \bottomrule
        \end{tabular}
        \label{tab:Z4}
        \end{table}

The anyons of the $\Z_4$ order are generated by a combinations of $\Z_4$ gauge charges $\e$ and gauges flux $\m$. These are each order-four, i.e. $\e^4=1=\m^4$. They are self bosons and have mutual $i$ statistics. The anyon types and their symmetry properties of the $\Z_4$ order are shown in Table~\ref{tab:Z4}. The $\e$ particle carries electric charge $e/2$ and no spin, and the $\m$ particle carries spin $\hbar/4$ and no charge. Time reversal symmetry maps $\m$ to $\m^{-1}=\m^3$, and $\m^2$ is a Kramers doublet, $\mathcal{T}^2(\m^2)=-\m^2$. 
A general anyon is a combination of $j$ $\e$ particles and $k$ $\m$ particles, which we abbreviate as: $(j,k) \equiv \e^j\m^k$. 
The content of the Abelian $\Z_4$ topological order is summarized in Table~\ref{tab:Z4}.

\subsection{Helical $U(1)_8$ FQSH order}
The $\ueight$ FQSH order can be viewed as two separate orders, one for each valley. The $U(1)_8$ order for the $K$-valley can be viewed as a $\nu=1/8$ bosonic Laughlin state for intra-valley Cooper pairs, described by a level-8 $U(1)$-Chern-Simons theory, and similarly for the $K'$-valley. The anyons of the $\ueight$ order can be labelled as $(n,m)$, with $n,m=0,1\cdots, 7$. The $(n,m)$ anyon carries charge $(n-m)e/4$ and spin $(n+m)\hbar/8$. Time reversal symmetry exchanges anyons in the two valleies, $\mathcal{T}:(n,m)\to (m,n)$. The anyon content and symmetry properties are shown in Table~\ref{tab:ueight}.
\begin{table}[h]
        \caption{{\bf $\ueight$ topological order: } $m,n\in \{0,1\cdots 7\}$. Total quantum dimension: $D=8$. Torus $\gsd=64$.\\
         }
        \renewcommand{\arraystretch}{1.3}
        \begin{tabular}{C{0.6in}C{0.2in}C{0.7in}C{0.6in}C{0.5in}C{0.5in}}
        \toprule
         Anyon ($a$) & $d$ & $\theta_a$ & charge ($e$) & $S^z$ ($\hbar$) & $\mathcal{T}(a)$ \\ 
        \midrule
        $(m,n)$ & 1 & $e^{i\pi/8(m^2-n^2)}$ & $\frac{m-n}{4}$ & $\frac{m+n}{8}$ & $(n,m)$  \\
        \bottomrule
        \end{tabular}
        \label{tab:ueight}
        \end{table}

\subsection{Non-Abelian FQSH states: Helical Pfaffian and variants}
While the $\Z_4$ FQSH has spin/valley entanglement, the remaining FQSH orders are all disentangled products of a topological order for spin-up electrons and its time-reversed conjugate for spin-down electrons, which we refer to as a helical version of the topological order.

Based on analogies between the \tbmo bands and Landau levels, a half-filled second band of \tbmo can be viewed as a helical version of a half-filled second Landau level.
The second Landau level for electrons in GaAs heterostructures is believed to form a non-Abelian topological order: the Pfaffian (Pf) state or its particle-hole symmetric cousin (PHPf)~\cite{wang2016half}.

\subsubsection{Chiral Pfaffian (Pf) FQH state}
Physically, the Pf state can be viewed as a topological superconductor of composite fermions (CFs): vortex-like objects formed from electrons ``attached" to $(-4\pi)$ flux~\cite{son2015composite}.
The elementary ($\pi$) superconducting vortex of the topological CF superconductor binds charge $1/4$ from the quantum Hall response, and also binds a Majorana zero mode of the Composite fermions, endowing it with non-Abelian properties.
The anyons of a Pf state can be labeled by an Ising anyon label, $\{1,\sigma,f\}$ where $f$ is the unpaired CF, and $\sigma$ is the non-Abelian vortex, and a vorticity $n=1,2,\dots 7,8\simeq 0$ with the constraint that $\sigma$ have odd vorticity and $1,f$ have even vorticity, and topological properties only depend on the vorticity modulo $8$. We use $a_n$ to denote an Pf anyon with Ising sector $a$ and vorticity $n$.

The anyon content almost has the same structure as a product of a non-Abelian Ising topological order and an Abelian $U(1)_8$ sector describing the fractional Hall response.
However, the resulting order differs from $\rm Ising \times U(1)_8$ in an important way.
Namely, the Pf state has a local fermion: the electron, whose topological superselection sector we denote by $c$. 
The anyon $f_4$ has precisely the same quantum numbers as the local electron, and should be identified with $c$.
Formally, one can achieve this by starting from ${\rm Ising}\times U(1)_8\times \{1,c\}$, and condensing $\<c^\dagger f_4\>\neq 0$, which identifies $f_4$ with the local electron, and also confines the 
${\rm Ising}\times U(1)_8$ excitations that braid non-trivially with $f_4$ such as $\sigma_{2n}$ and $1_{2n+1},f_{2n+1}$ thereby enforcing the constraint that  the Abelian (non-Abelian) particles respectively carry even (odd) vorticity.

\subsubsection{Chiral Particle Hole Pfaffian (PFPf) FQH State}

The PHPf state topological order simply corresponds to replacing the Ising sector in Pf with its time-reversed conjugate, $\overline{\rm Ising}$.
Physically, the PHPf and Pf states differ in the relative propagation direction of their neutral edge modes. In the Pf state, the neutral and charged edge modes are co-propagating giving thermal Hall conductance $\kappa^{xy} = \frac32$ (in units of that of a $\nu=1$ electronic integer quantum Hall edge).
For the PHPf, the neutral mode propagates in the opposite direction of the charged mode instead yielding $\kappa^{xy}=\frac12$.

This topological order was originally uncovered in the context of anomalous surface states of $3d$ electron topological insulators with a Dirac cone surface state~\cite{wang2013gapped,chen2014symmetry,metlitski2015symmetry}, and later gained theoretical~\cite{wang2016half,potter2017realizing} and possibly experimental relevance for the half-filled Landau level~\cite{zucker2016particle}. Adopting Son's modern viewpoint~\cite{son2015composite} of CFs as Dirac particles with an electric-dipole psuedospin-half degree of freedom, the PHPf simply corresponds to an s-wave pairing condensate of Dirac CFs, whereas the Pf state corresponds to a more complicated $d+id$ pairing of Dirac CFs~\cite{son2015composite}.

The $S$-matrix and fusion rules are the same as those for the Pf state listed above.

\subsubsection{Helical Pf and PHPf States}
We denote the helical Pfaffian state, consisting of a Pf state of spin-up and its time-reversed conjugate of spin-down electrons as: $\pf$. Similarly we define the helical PHPf as: $\phpf$. 
We denote the time-reversed anyons of the spin-down with an overbar, for example $\sigma_3\bar{1}_2$  denotes a composite of 3 non-Abelian vortices of spin-up and a double vortex of spin-down CFs. The fractional charge, spin, and exchange statistics of the excitations of the helical Pf and helical PHPf states are summarized in Table~\ref{tab:pf}.

We note that, while the chiral Pf an PHPf can be distinguished in thermal conductance measurements, their helical analogs both have vanishing thermal Hall conductance, and identical quantized two-terminal edge thermal conductance. Instead, these two orders differ only in the Abelian statistical phase factors of different anyon excitations, and can only be distinguished by more subtle interferometric measurements.

\begin{table}[t!]
        \caption{{\bf Helical Pfaffian ($\pf$) topological order: } $m,n\in \{0,1\cdots 7\}$. $a,b=1,\sigma,f$. $d_1=d_f=1,~d_\sigma=\sqrt{2}$. $\theta_f=-1,~\theta_\sigma=e^{\frac{i\pi}{8}}$. Total quantum dimension: $D=16$.
        The helical PHPf ($\phpf$) order has the same properties except with the replacement$\theta_\sigma\rightarrow e^{-i\pi/8}$.
        \\
         }
        \renewcommand{\arraystretch}{1.3}
        \begin{tabular}{C{0.6in}C{0.3in}C{0.8in}C{0.6in}C{0.4in}C{0.4in}}
        \toprule
         Anyon ($a$) & $d$ & $\theta_a$ & charge ($e$) & $S^z$ ($\hbar$) & $\mathcal{T}(a)$ \\ 
        \midrule
        $(a_m,\overline{b}_n)$ & $d_ad_b$ & $\theta_a\theta_b^*e^{\frac{i\pi(m^2-n^2)}{8}}$ & $\frac{m-n}{4}$ & $\frac{m+n}{8}$ & $(\overline{b}_n,a_m)$ \\
        \bottomrule
        \end{tabular}
        \label{tab:pf}
        \end{table}

\section{Edge states and Topological Defects}
\subsection{Generalities\label{app:Generalities}}
\subsubsection{Edge effective field theory}
A Euclidean spacetime field theory for gapless, symmetric edge field theory for the Abelian candidate states can all be written as two-component Luttinger liquids with Lagrangian density:
\begin{align}
\mathcal{L}_{\rm LL} = \frac{1}{4\pi} \sum_{I,J=1,2}\left[ ( i\d_\tau \phi_IK_{I,J} - v\d_x\phi_IG_{IJ}) \d_x\phi_J\right]\label{appeq:LL}
\end{align}
where $K$ is an integer matrix encoding the statistics of the $\phi$ fields. Specifically, excitations created by vertex operators $e^{i\ell_I\phi_I}$ (where $\ell_I$ are integer vectors) have commutation relations:
\begin{align}
e^{i\ell\cdot\phi(x)}e^{i\ell'\cdot \phi(y)} = e^{2\pi i \ell\cdot K^{-1}\cdot \ell' \Theta(x-y)}e^{i\ell'\cdot \phi(y)}e^{i\ell'\cdot \phi(x)} 
\end{align}
where $\Theta(x) = \begin{cases} 1 & x>0 \\ 0 & x<0 \end{cases}$ is the unit step function.
The velocity of the edge modes is $v$. The dimensionless matrix $G_{I,J}$ encodes the forward scattering interactions between different co-propagating edge modes, and sets the scaling dimension for vertex operators:
\begin{align}
\<e^{i\ell \cdot \phi(x)}e^{-i\ell\cdot \phi(y)}\> \sim \frac{1}{|x-y|^{2\Delta_\ell}},
\end{align}
Since the Luttinger liquid forms a critical phase with continuously evolving exponents, $\Delta_\ell$ depend continuously on the Luttinger parameters. Detailed expressions are given for specific examples below.

In addition, the non-Abelian orders have an additional helical, charge-neutral helical Majorana mode, $\chi_s$ where $s=\pm 1$ correspond to spin $\up$ or $\down$ respectively:
\begin{align}
\mathcal{L}_{\rm h-Majorana} = \sum_{I,J\in \{\up,\down\}} f_I (\zeta i\d_\tau - v\sigma^z_{IJ}\d_x)f_J
\label{appeq:majorana}
\end{align}
where $\zeta=\pm 1$ depending on the particular brand of non-Abelian order.
In each case, each spin Majorana mode is coupled to the Abelian sector and sees its corresponding superconducting vortex as a $\pi$ flux (this constraint is not explicitly captured in the free-fermion field theory description above, which should hence be viewed as somewhat schematic).
The gauged-Majorana fermion forms a helical Ising CFT corresponding to a chiral Ising CFT of right-moving fields with spin-up and valley-$K$, and a left moving chiral $\overline{\rm Ising}$ CFT of left-moving fields with spin-down and valley $(-K)$.

\subsubsection{Gapped boundaries and anyon condensation}
The gapped boundaries of non-chiral topological orders can be equivalently classified purely from the data of the topological order. Each gapped boundary corresponds to condensing a Lagrangian subset, $\mathcal{A}$ of anyons defined by the properties that each of the condensed anyons is a boson, $\theta_a=1\forall a\in \mathcal{A}$ (for non-Abelian particles, condensation further requires that one of the fusion channels for $a\times b = \sum_c N_{ab}^c c$ contains at least one $c\in \mathcal{A}$, $\forall a,b\in \mathcal{A}$), and that every anyon in the topological order is either condensed, or braids non-trivially one one of the condensed anyons (and hence is confined by the condensate).

In the field theory description, these boundary condensations are implemented by adding \emph{local} terms that pin the boundary fields that insert the condensed anyons, as we explain for each example in-turn below.

\subsection{Spin-valley-entangled $\Z_4$ order}
\subsubsection{Edge States}
As described in the main text, the edge field theory for the $\Z_4$ FQSH consists of a non-chiral Luttinger liquid (Eq.~\ref{eq: HLL}) with $I,J\in \{\e,\m\}$, and $e^{i\phi_{\e,\m}}$ respectively creating $\e$ or $\m$ particles at the edge (there is no neutral Majorana sector for Abelian topological orders). The K-matrix: $K_{IJ}=4\sigma^x_{IJ}$ where $\sigma^x$ is the standard $X$-Pauli matrix, encodes the mutual statistics between $e$ and $m$: $\theta_{e,m} = e^{2\pi i/4}$. 

The non-trivial symmetry action on the edge fields are:
\begin{align}
U(1)_c: &\phi_e \rightarrow \phi_e+\alpha/2, \nonumber\\
U(1)_{\rm sv}: &\phi_m \rightarrow \phi_m + \alpha/4, \nonumber\\
\mathcal{T}: &\phi_e\rightarrow -\phi_e, \nonumber\\
\mathcal{T}:& \phi_\m \rightarrow \phi_\m +\pi/4,
\end{align}
(accounting for the anti-unitary nature of $\mathcal{T}:i\rightarrow -i$ the last line ensures that $e$ is $\mathcal{T}$-symmetric but $\mathcal{T}:\m\mapsto \m^{-1}$).

The commutation relations: $[\phi_I(x),\d_y\phi_J(y)] = 2\pi iK^{-1}_{IJ}\delta(x-y)$ and charge assignments above, imply that the electrical charge and charge density, $Q,\rho(x)$, and spin/valley charge density $S^z,\rho_{\rm sv}$ operators are:
\begin{align}
Q &= \int dx \rho(x), &\rho(x) = \frac{1}{\pi}\d_x\phi_m \nonumber\\
S^z &= \frac12\int dx \rho_{\rm sv}(x),&\rho_{\rm sv}(x) = \frac{1}{\pi} \d_x\phi_e
\end{align}
(note the conventional factor of $\frac 12$ in the spin/valley charge definition makes spin/valley ``charge" quantized in the same units of $\frac12$ as the electron spin).
 The $U(1)_{\rm c/sv}$ symmetry generators are respectively: $U_{\rm c}(\alpha) = e^{-i\alpha Q}$, $U_{\rm sv}(\alpha) = e^{-i\alpha S^z}$.

The most general interaction parameter matrix consistent with these symmetries can be parameterized by a single dimensionless coupling constant $g$ as: $G_{IJ} = \begin{pmatrix} 1-g& 0 \\ 0& 1+g \end{pmatrix} = \delta_{IJ}-g\sigma^z_{IJ}$. Off-diagonal couplings that appear with terms $\sim \d_x\phi_e\d_x\phi_m$ are forbidden by time-reversal symmetry. 
Since charge density is proportional to $\d_x\phi_m$, we see that $g>0$ corresponds to repulsive interactions.
The resulting scaling dimension for vertex operator $e^{i\ell\cdot \phi}$ are:
\begin{align}
\Delta_{\ell} = \frac{1}{8}\left(\ell_\e^2 \sqrt{\frac{1+g}{1-g}}+\ell_\m^2 \sqrt{\frac{1-g}{1+g}}\right).
\end{align}
Terms such as $-\int dx \cos(\ell\cdot \phi)$ are relevant when $\Delta_{\ell}<2$.
From the above expression, we see that repulsive interactions ($g>0$) enhance the spin correlations, whereas attractive interactions ($g<0$) enhance the charge correlations.

\subsubsection{Gapped Boundaries}
The gaplessness of the Luttinger-liquid edge modes described above is protected by symmetry.
Breaking the protecting symmetry can open a gap in the edge. 
Breaking $U(1)_c$ via a superconducting (SC) proximity effect is described in the main text. 
In addition, there are two alternative gapped boundaries arising from breaking $U(1)_{\rm sv}$ and $\Z_2^T$ via an inter-valley coherent (IVC) spin-density wave (SDW) order or breaking all the symmetries via coexisting SC and IVC orders.

In the following, it is important to bear in mind that, while the the $z$-component of the spin-valley degree of freedom is equivalent to a $z$-component of spin, an in-plane spin-valley polarization has no net $xy$ spin moment, and does not couple to an in-plane magnetic field, but rather exhibits only an in-plane SDW pattern with wave-vector $2K$.

\paragraph{IVC/SDW boundary}
Inducing an inter-valley coherent (IVC) spin density wave order (SDW) order with an in-plane spin texture that is modulate at wave-vector $2K$ at the edge can effectively condense $\m$ anyons, resulting in a magnetic edge:
\begin{align}
H_M &= \frac{\Delta_{\rm IVC}}{2} \int dx e^{i(2K_xx-\varphi)} c^\dagger_{\up,K}(x)c^{\vphantom\dagger}_{\down,-K}(x)
\nonumber\\
&\approx -\lambda_M\int dx \cos [4\phi_m(x)-\varphi+2k_Fx ].
\end{align} 
where $K_x$ is the projection of the $K$-point of the TMD Brillouin zone onto the direction parallel to the edge, and $k_F$ is the Fermi wave-vector of the edge.

If the chemical potential of the edge is tuned to $k_F=0$, then this term has scaling dimension $\Delta_{\ell=(0,4)} = 2\sqrt{\frac{1-g}{1+g}}$ and is perturbatively relevant when $g < 1$, i.e. when $g\geq 3/5$ (note we also require $-1<g<1$ for stability of the edge), or can become non-perturbatively relevant if $\lambda_M$ is sufficiently strong.

When relevant, this term locks the phase of the edge mode to: $\phi_m = \frac{\varphi_B}{4} +\frac{2\pi s}{4}$, which has four degenerate minima: $s=0,1,2,3$. Physically, these minima  correspond to different total fractional value of the spin, $S^z=j/4$, on the edge. 
The edge degeneracy is only present when there are multiple edges, since there is a global constraint that the system's total charge and spin, $(Q,S^Z)$ must correspond to those of integer numbers, $N_{\up,\down}$, of up and down spin electrons, $(Q,S^z) = (N_\up+N_\down,\frac12 N_{\up}-\frac12 N_\down)$. Since the fractional spin of the edge comes along with no charge, this constraint implies that the sum of $j$ for all the edges in the system must be a multiple of four. Namely, given $N$ distinct edges with the same magnetic gap, there are $4^{N-1}$ distinct minima, accounting for this single global constraint.
For edges of total length $\ell$, the splitting of these between these degenerate minima occurs through instanton tunneling events which have action cost $\sim e^{-c\lambda_M\ell}$ where $c$ is a constant that can be approximated by standard instanton/WKB techniques.

Alternatively, and as discussed in~\cite{chou2024composite}, short range disorder breaks translation symmetry, and can induce glassy order, which locks $\phi_m(x) = f(x) + \pi s/4$ to a random component of disorder. Here, the edge is gapless but localized, and there remains a global fourfold topological degeneracy that could be used as a topological memory.

\paragraph{Anyon condensation picture}
The gapped boundaries can also be simply understood from the anyon condensation perspective.
As described in the main text, the superconducting boundary corresponds to pinning $\phi_e$, which loosely-speaking gives an expectation value to the operator $e^{i\phi_e}$ that inserts an $e$ particle at the edge. Hence, we can view this edge as hosting an $\e$ particle condensate. Since condensing $\e$ also condenses composites $\e^j$. The corresponding condensation algebra is denoted: $\mathcal{A}^\e = \sum_{j=0}^3\e^j$ (with $\e^0\equiv 1$).
Due to the fractional charge of $\e$, the $\e$ condensation necessarily breaks charge conservation symmetry, which can be diagnosed by long-range order in the local order parameter $e^{4i\phi_e}$.
Similarly, the IVC/SDW boundary can be obtained by condensing $\m$ particles at the edge, with corresponding condensation algebra $\mathcal{A}^{\m}=\sum_{j=0}^{3} \m^j$. The IVC/SDW boundary has local IVC order parameter $e^{4i\phi_\m}$, and breaks both $U(1)_{\rm sv}$ and $\Z_2^T$ symmetries, but preserves a (non-Kramers) anti-unitary symmetry: $\tilde{\mathcal{T}}=e^{i\pi S^z}\mathcal{T}$.

In addition, there is a third possible boundary corresponding to simultaneously condensing $\e^2$ and $\m^2$: $\mathcal{A}^{(\e^2,\m^2)} = (1+\e^2)(1+\m^2)$, which breaks all three protecting symmetries, and corresponds to a combination of SC and IVC/SDW order. This boundary is less simple to describe in the bosonized field theory, since it does not simply correspond to pinning the $\phi_{e,m}$ fields. 
Rather, it corresponds to pinning $\phi_e(x) = \frac{\pi q}{2}+\pi \alpha(x)$ and $\phi_m(x) = \frac{\pi s}{2}+\pi\beta(x)$ with $\alpha,\beta \in \{0,1\}$ being Ising fields, which are disordered such that $e^{2i\phi_{e,m}}$ has an expectation value, but neither $e^{i\phi_{e,m}}$ are condensed.


\subsubsection{Twist defects}
\begin{figure}[t!]
\includegraphics[width = 0.4\textwidth]{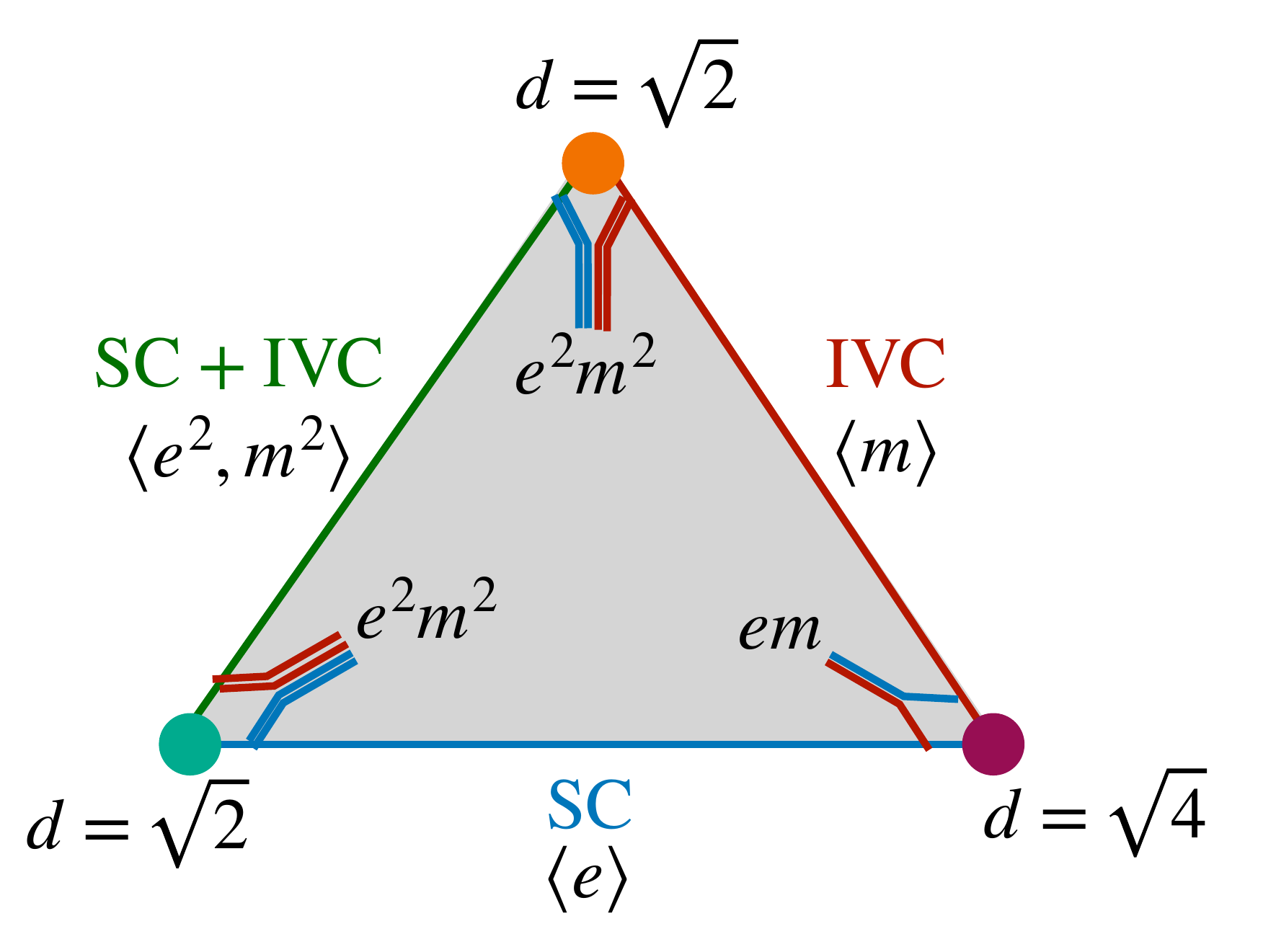}
\caption{{\bf Twist defects  -- } Three different types of boundaries (colored lines) and twist defects (dots at corners). $\<\dots\>$ list the anyons condensed at each boundary. The twist defects are either Majorana bound states with quantum dimension $d=\sqrt{2}$ or $\Z_4$ parafermions with quantum dimension $d=\sqrt{4}$. The logical operators for each defect correspond to anyon lines that can be absorbed near each defect, as indicated by the labeled lines terminating in each corner. Note, that due to the overall constraint that the total topological charge of the system is $1$, multiple twist defects of a given type are generally required to achieve a nontrivial topological ground-space. While the anyon labeling here is shown for the $\Z_4$ FQSH order, the gapped boundaries and twist defects of the other candidate orders can all be mapped onto the $\Z_4$ example as described in the text.
}
\label{appfig:twist}
\end{figure}
As previously explained in a number of works, reviewed in~\cite{alicea2016topological}, at the junction between a superconducting ($\e$-condensed) boundary and an IVC ($\m$-condensed) boundary, there is a point-like twist defect denoted by $\epsilon$, which has the property that braiding an $\e$ particle around the twist defect changes (``twists") it into an $\m$ particle. 
This twist defect is a non-Abelian $\Z_4$ parafermion mode, a $\Z_4$ generalization of the Majorana fermion, with quantum dimension $d_\epsilon = \sqrt{4}=2$, and fusion rules:
\begin{align}
        \epsilon\times \epsilon=1+em^3+e^2m^2+e^3m.
\end{align}


On the other hand, the junction between a all-symmetry-breaking boundary($(e^2,m^2)$ condensed) and an $e$- or $m$- condensed boundary, there is a Majorana fermion defect, denoted as $\epsilon_e$ and $\epsilon_m$ respectively, with quantum dimension $\sqrt{2}$ and fusion rules 
\begin{align}
    \epsilon_e\times \epsilon_e=1+em^2,~\epsilon_m\times \epsilon_m=1+e^2m
\end{align}
The three types of defects and the logical operators associated with the degenerated ground spaces are shown in Fig.~\ref{appfig:twist}.


\subsection{Gapped boundaries of helical FQSH orders}
In this section, we describe general features of gapped boundaries and Cheshire qudits made from helical FQSH orders.

Denote the anyons of a helical topological order by $(a,\bar{b})$ where $a$ ($\bar{b}$) label anyons in the chiral (anti-chiral) topological order of spin-up (down) electrons respectively.

\begin{figure}
    \vspace*{-0.5cm}
    \hspace*{-0.5cm}
    \includegraphics[width=0.65\textwidth]{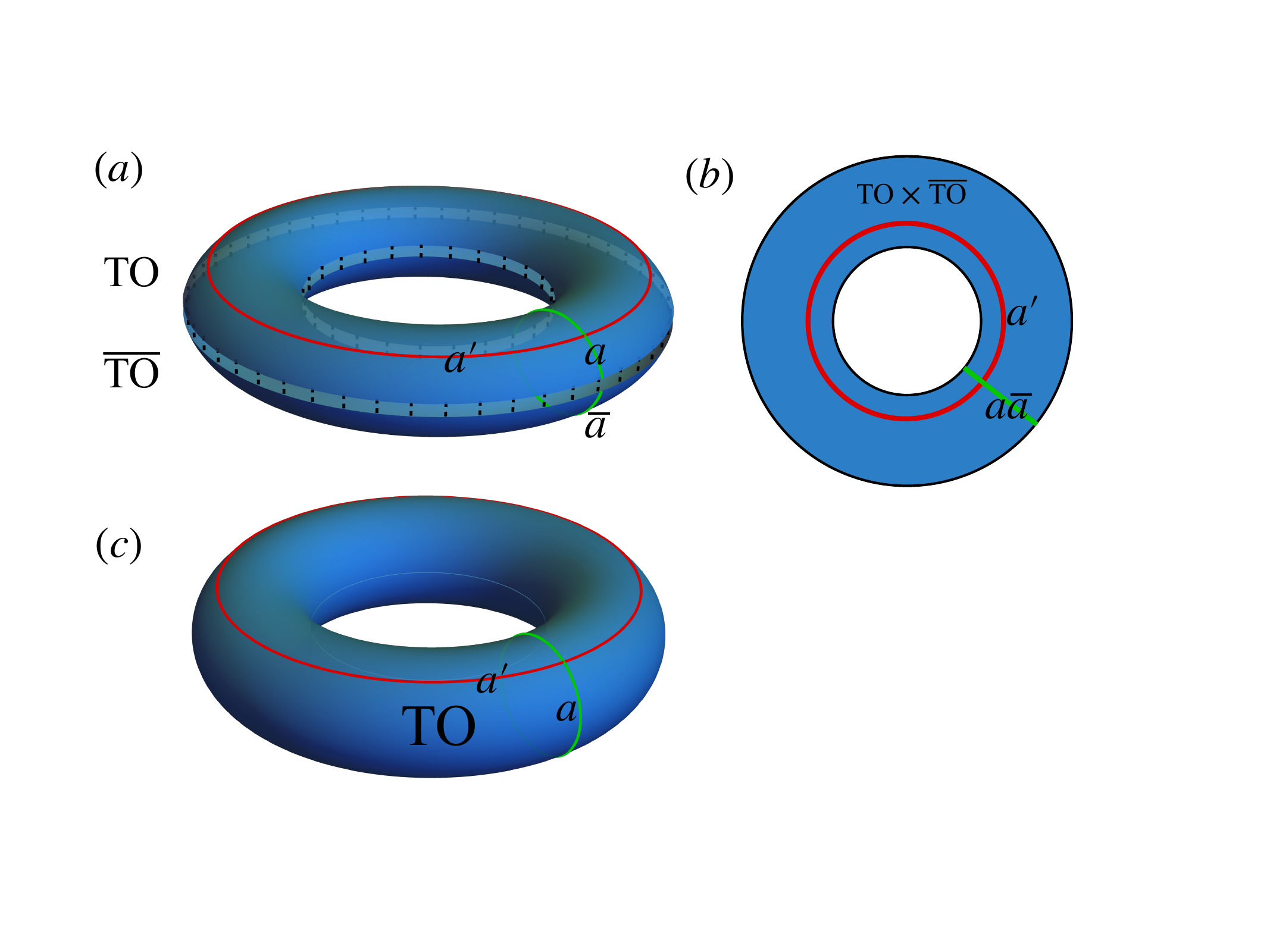}
    \vspace*{-1.5cm}
    \caption{A helical FQSH state can be viewed as a sheet of a chiral topological order (TO) stacked with a sheet of its time-reversal ($\overline{\text{TO}}$). (a) The helical FQSH state is put on a annulus geometry, with inner and outer boundaries gapped out by condensing the pairs $a\overline{a}$, indicated by the vertical dashed lines. This identifies the $a$-anyon of the top sheet with the $\overline{a}$-anyon of the bottom sheet. The green anyon lines consist of an $a$ line on the top sheet and an $\overline{a}$ line on the bottom sheet. Due to the condensation of $a\overline{a}$, the two green lines can meet at the inner and outer boundaries of the annulus without creating excitations. The red anyon line is an $a'$ line in the top sheet.  (b). The top view, showing a $\text{TO}\times \overline{\text{TO}}$ living in a annulus with inner and outer boundaries having $a\overline{a}$ condensed. (c) The condensation $\<a\overline{a}\>$ at the boundaries effectively sews the top sheet with the bottom sheet, leading to an equivalent description as a single TO on a torus. The red and green anyon lines now live on different non-contractible cycles of the torus. }
    \label{appfig:torus}
\end{figure}

For helical topological orders there is a simple relation between i) the properties of an array of Cheshire qudits made from a sheet with $N$ trivial holes punched out and all boundaries gapped by condensing a Lagrangian set of anyons, and ii) the state of the chiral topological order for spin-up particles defined on a genus-$N$ surface (See Fig.~\ref{appfig:torus} for illustration of $N=1$)

Gapped boundaries of a helical FQSH order correspond to condensing a Lagrangian set of $(a,\bar{b})$. Since the underlying chiral (antichiral) orders for each spin-species separately have ungappable edges, it is impossible to gap the edge by condensing particles such as $(a,1)$ or $(1,\bar{a})$. 

Based on simple symmetry considerations, there is always a superconducting boundary obtained by condensing $\{(a,\bar{a}^{-1})\}$ for all $a$, where $a^{-1}$ denotes the antiparticle of $a$. Since particle/anti-particle conjugation preserves the statistical properties, and flips the electrical- and spin/valley- charges, this set of anyons are spinless bosons, and at least one carries a non-zero electrical charge (hence condensing it leads to a superconductor). Similarly, there is also a $U(1)_c$-symmetric, $U(1)_{\rm{sv}}$ and $\Z_2^T$ breaking magnetic/intervallley coherent (IVC) edge with spin-density wave order obtained by condensing the charge-neutral $\{(a,\bar{a})\}$.

For both of these simple superconducting and IVC boundaries, an array of Cheshire qudits obtained by punching out $N$ holes in a helical FQSH sheet, can be equivalently viewed as the chiral TO for spin-up particles on a closed genus-$N$ surface.
Namely, viewing the chiral TO of up (down) electrons as the ``top" (``bottom") sheet of a higher genus surface, and the boundary anyon condensation as gluing these sheets together at their edges if we identify $a$ on the top surface by $\bar{a}$ or $\bar{a}^{-1}$ on the bottom surface (see Fig.~\ref{appfig:torus}).

This perspective enables one to simply compute the GSD, and logical operators of Cheshire qudits in the helical FQSH order from knowledge of those of the chiral TO on a closed torus.
Namely, the logical operators of each Cheshire qudit, map onto anyon loop operators $W_a^i$ that drag an $a$ particle around the $i^{\rm th}$ cycle of the closed genus-$N$ surface.

For example for single Cheshire qudit, made from an annulus of helical FQSH with gapped boundaries, which maps to a chiral FQH order on a torus, there are two inequivalent cycles, which we label as $x,y$.
The topological ground-states can be generated from a single reference state, $|1\>$ that is an eigenstate of $W^x_1$, defined through $W^x_a|1\>=|a\>$, where $a$ runs over the possible anyon types in the chiral FQH order.
The algebra of logical operators is then defined by:
\begin{align}
    W^x_b|a\> &= \sum_c N^c_{ab}|c\> \nonumber\\
    W^y_b|a\> &= \frac{S_{ab}}{S_{1b}}|a\>
    \label{appeq:Wops}
\end{align}
where $N^c_{ab}$ are the fusion coefficients (with associated fusion rules: $a\times b = \sum_c N^c_{ab}$), and $S_{ab}$ is the modular $S$-matrix of the FQH order.

For an Abelian \TO, there is always a unique fusion channel, and $\frac{S_{ab}}{S_{1b}}=\theta_{a,b}$ is simply the exchange phase for anyons $a,b$, so the operators $W^i_a$ are unitary, and can (in principle) be implemented by creating an $a$-anyon/anti-anyon pair, dragging $a$ around cycle $i$, and then re-annihilating it with its antiparticle partner.

For a non-Abelian order, the transport around cycle $i$ can change the fusion channel of the anyons so that it may not necessarily be able to re-annihilate with its antiparticle partner with 100\% probability. Hence, some of the $W^i_a$ operators are inevitably non-unitary, and can only be implemented by a combination of unitary dynamics, measurement, and feedback.

\subsubsection{Inter-edge tunneling for superconducting Cheshire qudits}
The mapping from Cheshire qudits of helical \TO's to chiral \TO on a higher-genus surface is useful for analyzing the inter-edge tunneling.
Consider the simplest setup of an annular Cheshire qudit made form a helical FQSH state, with superconducting boundaries, which can be mapped onto a chiral \TO on a torus.

The general inter-edge tunneling Hamiltonian can be constructed from the anyon-tunneling operators in (\ref{appeq:Wops}):
\begin{align}
    H_{\Gamma} = -\sum_a \Gamma_a e^{i\theta q_a/2}W^y_a+{\rm h.c.}
\end{align}
where $\Gamma_a$ is the inter-edge tunneling amplitude for anyons of type $a$, $\theta$ is the phase difference between the superconductors in the inner and outer edges, and $q_a$ is the charge of anyon $a$.

The corresponding supercurrent operator:
\begin{align}
    I_S = -\frac{2\pi}{\Phi_0}\frac{\d H_\Gamma}{\d \theta} = \frac{2\pi}{\Phi_0}\sum_{a,b} q_b 
    {\rm Re}\left[\Gamma_b \frac{S_{a,b}}{S_{b,1}} e^{i\theta q_a/2}\right]|a\>\<a|.
\end{align}
While this expression depends on generally unknown amplitude and phases of different tunneling amplitudes, $\Gamma_a$, for generic (non-fine-tuned) values of $\Gamma_a$ and $\theta$, there will be a unique value of $I_S$ for each state $|a\>$, and a properly calibrated supercurrent measurement can be used to readout the state of the Cheshire qudit.
Similarly, absent accidental degeneracies, there will be a unique ground-state of $H_\Gamma$, enabling one to remove the GSD by gate-controlled inter-edge tunneling, and detect the associated change in the thermal entropy as described for the $\Z_4$ FQSH order in the main text.

We note for future reference, the modular $S$-matrix of the chiral Pf and PHPf orders are~\cite{bernevig2017topological}:
\begin{align}
    S^{\rm (Pf)}_{a_n,b_m} =\frac{e^{2\pi i nm/8}}{4\sqrt{2}}\begin{pmatrix}
        1 &\sqrt{2}&1\\
        \sqrt{2}&0&-\sqrt{2}\\
        1&-\sqrt{2}&1
    \end{pmatrix},
\end{align}
and its fusion rules are: $\sigma_n\times\sigma_m = 1_{n+m}+f_{n+m}$, $a_n\times 1_m = a_{n+m}$, $\sigma_n\times f_m = \sigma_{n+m}$, i.e. the following fusion coefficients are equal to one: $N^{a_n}_{a_n,1_m}\forall a,N^{1_{n+m}}_{\sigma_n,\sigma_m},N^{f_{n+m}}_{\sigma_n,\sigma_m},,N^{1_{n+m}}_{f_n,f_m},N^{\sigma_{n+m}}_{\sigma_{n},f_m}$ and the remainder vanish.

For the Abelian $U(1)_8$ chiral order, the $S$ matrix is simply $S_{n,m} = \frac{1}{8}e^{2\pi i nm/8}$, and the fusion rules are $N^{k}_{n,m} = \delta_{k,(n+m)\mod 8}$.

\subsection{Helical Pfaffian order}

\subsubsection{Edge states}
The helical Pfaffian order has edge states consisting of both a charged Luttinger liquid, (\ref{appeq:LL})
with $K_{I,J}=8\sigma^z_{I,J}$ where $I,J\in \{\up,\down\}$, and a co-propagating neutral Majorana fermion mode, (\ref{appeq:majorana}) with $\zeta-+1$. 
Here, the vertex operator $e^{i(n\phi_\up(x)+m\phi_\down(x))}$ creates the edge avatar of the bulk $1_n\bar{1}_m$ particle, and $f_{\up},f_{\down}$ are respectively the edge avatar of the bulk $f,\bar{f}$ particles.
The edge avatar of the bulk $\sigma_{n}$ excitation is a twist defect (disorder operator) for the Majorana fermions, and inserting this excitation at position $x$ induces a phase branch cut in the Majorana fermion field: $f(x')\rightarrow (-1)^{\theta(x-x')}f(x)$.
Implicit in this description, the edge also contains a gapped local spin-1/2 fermion (the electron), $c_\sigma$, which should be identified with $f_4$, i.e. the objects $e^{4\sigma i\phi_\sigma}f_\sigma c^\dagger_\sigma$ are condensed, which enforces the constraint that $\sigma$ excitations come with odd vorticity $n$.

While the description of gapped edges can, in principle be carried out in this field theory, we find it more convenient to analyze the gapped, symmetry-breaking non-Abelian edges and resulting Cheshire qudit properties through the algebraic anyon description and via the corresponding tunneling operators in (\ref{appeq:Wops}).

\subsubsection{Gapped boundaries}
It turns out that all the gapped boundaries of the helical Pfaffian \TO are in one-to-one correspondence with those of the $\Z_4$ \TO described above. To see this, we first show that any gapped boundary necessarily requires condensing the anyon $1_4\bar{1}_4$. Second we show that condensation of $1_4\bar{1}_4$ reduces the helical Pfaffian order to the $\Z_4$ order (near the edge where the condensation occurs), in the sense that all the remaining anyons(un-condensed and deconfined anyons) at the boundary form a $\Z_4$ order after $1_4\bar{1}_4$ condensation.
Then, we can simply use the dictionary between reduced-helical Pfaffian and $\Z_4$ orders to identify the analogs of the $e$-condensed, $m$-condensed and $(e^2,m^2)$-condensed boundaries of the $\Z_4$ order. Importantly, we note that, while there is a correspondence between the \emph{number} of gapped boundaries, the resulting Cheshire qudits will differ from those of the $\Z_4$ order in their GSD and logical operations as we describe in detail below.

\paragraph{Aside: topological orders with local fermions}
The presence of a local fermion, the unfractionalized electron, that plays a crucial role in the helical Pfaffian order (the $f_4$ and $\bar{f}_4$ ``anyons" are identified with the spin-up and down electrons respectively) leads to some standard, but technical subtleties~\cite{aasen2019fermion,lou2021dummy,zhang2024hierarchy}.
As reviewed in~\ref{app:Generalities}, gapped boundaries of non-chiral topological orders(\TO) are classified by Lagrangian algebras, which physically describe condensation of anyons of the bulk \TO such that all anyons are trivialized by the condensation. We note that in a fermionic \TO it is possible to condense bound state of fermionic anyons and physical fermions, and such condensations are controlled by the so-called fermionic condensable algebras. An alternative approach is to consider the minimal modular extension of the fermionic \TO, which is a bosonic \TO, described by a standard modular tensor category. The modular extension can be viewed as obtained by gauging fermion parity of the fermionic \TO. The original fermionic \TO can be recovered from the modular extension by condensing $fc$, where $f$ is a fermionic anyon of the modular extension, and $c$ is the physical fermion. In our case, a modular extension of the $\pf$ order is $\text{Ising}\times U(1)_8\times \overline{\text{Ising}}\times U(1)_{-8}$, and $\pf$ is recovered from the modular extension by condensing $f_4c, \overline{f}_4c$. Notice this also implies the boson $f_4\overline{f}_4$ is condensed in $\pf$.

\paragraph{Necessity of condensing $1_4\overline{1}_4$}
We first show that any Lagrangian condensation of $\pf$ necessarily condenses $1_4\overline{1}_4$. The topological spins of anyons in the spin-up and spin-down sectors are listed below:
\begin{align}
    \theta_{1_n} &= e^{i\pi n^2/8} = (-1)\theta_{f_n}\nonumber\\
    \theta_{\sigma_n} &= e^{i\pi(1+n^2)/8}
    \nonumber\\
    \theta_{\bar{a}_n} &= \theta_{a_n}^*
\end{align}
Firstly, if any non-Abelian anyon is condensed, it must take the form $\sigma_n\overline{\sigma}_m$, since the quantum spin of a single $\sigma_{n}/\overline{\sigma}_m$ anyon can not be trivialized by combining with any Abelian anyons. Notice that we have
\begin{align}
   ( \sigma_n\overline{\sigma}_m)^2=1_{2n}\overline{1}_{2m}(1+f+\overline{f}+f\overline{f})
\end{align}
If $\sigma_n\overline{\sigma}_m$  is condensed, then one of the fusion outcomes above must also condense. Since all fusion outcomes above square to $1_{4n}\overline{1}_{4m}$, we conclude that $1_{4n}\overline{1}_{4m}$ is condensed. For $n,m$ odd we have $4n\equiv 4m\equiv 4 \mod 8$. Thus $1_4\overline{1}_4$ is condensed. 

Next consider the case where only Abelian anyons are condensed. The only bosonic anyons that are not roots of $1_4\overline{1}_4$ are $1_4, \overline{1}_{4}, f_4\overline{f},f\overline{f}_4, f\overline{f}$. However, these five anyons are mutually bosonic. Thus in order to have a Lagrangian condensation, they must be condensed simultaneously, which implies condensation of $1_4\overline{1}_4$. This completes the proof that any Lagrangian condensation of $\pf$ must condense $1_4\overline{1}_4$.

\paragraph{Reduction of helical Pfaffian order by $(1_4,\bar{1}_4)$ condensation}
We next show that condensing the anyon $(1_4,\overline{1}_4)$ reduces $\pf$ to the $\Z_4$ order $D(\Z_4)$. Since $f_4\overline{f}_4$ is always condensed in the $\pf$ order, condensing $1_4\overline{1}_4$ also condenses $f\overline{f}$. It is well-known that condensing $f\overline{f}$ in $\text{Ising}\times \overline{\text{Ising}}$ reduces it to $D(\Z_2)$. In particular, the non-Abelian anyon $\sigma \overline{\sigma}$ splits into $\widetilde{e}+\widetilde{m}$ of a $\Z_2$ toric code, and $f,\overline{f}$ both become the $\widetilde{f}=\widetilde{e}\times \widetilde{m}$ anyon of the $\Z_2$ toric code . On the other hand, condensing $1_4\overline{1}_4$ in the $U(1)_8\times U(1)_{-8}$ order reduces it to $D(\Z_4)$, generated by the deconfined anyons $e':=1_1\overline{1}_1$ and $m':=1_1\overline{1}_{-1}$. Therefore, after condensing $1_4\overline{1}_4$ and $f\overline{f}$, the modular extension $\text{Ising}\times U(1)_8\times \overline{\text{Ising}}\times U(1)_{-8}$ is reduced to $D(\Z_2)\times D(\Z_4)$. To obtain a fermionic \TO, we need to further condense the bound states $f_4c\sim \overline{f}_4c= \widetilde{f}\times e'^2m'^2c$. This reduces $D(\Z_2)\times D(\Z_4)$ to a single $D(\Z_4)$, generated by $\tilde{e}e'$ and $\tilde{e}m'$, stacked with a trivial fermionic \TO $\{1,c\}$.

\paragraph{Classification of gapped boundaries of the helical Pfaffian order}
By the previous discussion, condensing $1_4\overline{1}_4$ reduces the $\pf$ order to the $\Z_4$ order. Therefore there are as many gapped boundaries of the $\pf$ order as there are of the $\Z_4$ order.  Due to the relation to gapped boundaries of the $\Z_4$ order, we will call the three gapped boundaries of the $\pf$ order the $e$-, $m$- and $(e^2,m^2)$-boundaries. For completeness, we list the corresponding condensable algebras in $\pf$ below. 
\begin{align}
    \mathcal{A}^e=&1+1_2\overline{1}_2+1_4\overline{1}_4+1_6\overline{1}_6\nonumber\\
    &+f\overline{f}+f_2\overline{f}_2+f_4\overline{f}_4+f_6\overline{f}_6\nonumber\\
    &+\sigma_1\overline{\sigma_1}+\sigma_3\overline{\sigma_3}+\sigma_5\overline{\sigma_5}+\sigma_7\overline{\sigma_7}\\
     \mathcal{A}^m=&1+1_2\overline{1}_6+1_4\overline{1}_4+1_6\overline{1}_2\nonumber\\
    &+f\overline{f}+f_2\overline{f}_6+f_4\overline{f}_4+f_6\overline{f}_2\nonumber\\
    &+\sigma_1\overline{\sigma_7}+\sigma_3\overline{\sigma_5}+\sigma_5\overline{\sigma_3}+\sigma_7\overline{\sigma_1}\\
    \mathcal{A}^{(e^2,m^2)}=&1+1_2\overline{1}_2+1_4\overline{1}_4+1_6\overline{1}_6+1_4\nonumber\\
    &+\overline{1}_4+1_2\overline{1}_6+1_6\overline{1}_2
\end{align}
Notice that these three condensations are not Lagrangian when viewed as condensations in the modular extension $\text{Ising}\times U(1)_8\times \overline{\text{Ising}}\times U(1)_{-8}$. For example, for $\mathcal{A}^e$, the deconfined anyons are generated by $f_4$ and $1_1\overline{1}_{1}$. But $f_4$ is identified with the physical fermion in $\pf$, and $1_1\overline{1}_1$ is confined by the condensation $f_4c$. Therefore $\mathcal{A}^e$ is a Lagrangian condensation of $\pf$. Similarly, the condensations $\mathcal{A}^m,~\mathcal{A}^{e^2,m^2}$ are also Lagrangian in $\pf$.

\subsubsection{Twist defects}
Since the gapped boundaries of the $\pf$ order can be viewed as gapped boundaries of the $\Z_4$ order, the twist defects between different boundaries have the same properties as those of the $\Z_4$ order, namely the defect between an $e$-boundary and an $m$-boundary is a $\Z_4$-parafermion, and the defect between an $e/m$ boundary and an $(e^2,m^2)$ boundary is a Majorana fermion.

\subsection{Helical PH-Pfaffian order}
\subsubsection{Edge states}
The edge states of the helical PHPf order are nearly identical to those discussed above for the helical Pf order, except the neutral Majorana sector has opposite helicity, i.e. is described by (\ref{appeq:majorana}) with $\zeta=-1$.
Again, rather than using this explicit field theory, we find it more convenient to analyze the gapped boundaries and resulting Cheshire qudits through the algebraic anyon framework.

\subsubsection{Gapped boundaries}
Following similar arguments as for the $\pf$ order, it is readily seen that  any gapped boundary of the $\phpf$ order must condense the anyon $\overline{1}_41_4$. Similarly, condensing $\overline{1}_41_4$ in the $\phpf$ order also reduces the topological order to a $\Z_4$ order. Thus we have again three types of gapped boundaries, corresponding to condensing $e$, $m$ and $(e^2,m^2)$ of the $\Z_4$ order. The explicit condensable algebras are listed below. 
\begin{align}
    \mathcal{A}^e=&1+\overline{1}_21_2+\overline{1}_41_4+\overline{1}_61_6\nonumber\\
    &+\overline{f}f+\overline{f}_2f_2+\overline{f}_4f_4+\overline{f}_6f_6\nonumber\\
    &+\overline{\sigma_1}\sigma_1+\overline{\sigma_3}\sigma_3+\overline{\sigma_5}\sigma_5+\overline{\sigma_7}\sigma_7\\
     \mathcal{A}^m=&1+\overline{1}_21_6+\overline{1}_41_4+\overline{1}_61_2\nonumber\\
    &+\overline{f}f+\overline{f}_2f_6+\overline{f}_4f_4+\overline{f}_6f_2\nonumber\\
    &+\overline{\sigma_1}\sigma_7+\overline{\sigma_3}\sigma_5+\overline{\sigma_5}\sigma_3+\overline{\sigma_7}\sigma_1\\
    \mathcal{A}^{(e^2,m^2)}=&1+\overline{1}_21_2+\overline{1}_41_4+\overline{1}_61_6+1_4\nonumber\\
    &+\overline{1}_4+\overline{1}_21_6+\overline{1}_61_2
\end{align}
\subsubsection{Twist defects}
Similar to the $\pf$ order, the defect between an $e$-boundary and an $m$-boundary is a $\Z_4$ parafermion, and the defect between an $e/m$-boundary and an $(e^2,m^2)$-boundary is a Majorana fermion. 

\subsection{Edge states of the helical $U(1)_8$ order}

\subsubsection{Gapped boundaries: field theory}
The edge states of the helical $U(1)_8$ order are described by the Luttinger liquid action (\ref{appeq:LL}), with flavor labels $I,J\in \{\up,\down\}$, and $K=8\sigma^z$. In this notation $e^{i\phi_{\up}}$ creates a right moving quasiparticle with charge $1/4$ and  $S^z=1/8$, and $e^{i\phi_\down}$ creates a left moving quasihole with charge $-1/4$ and spin $S^z=1/8$.

The non-trivial symmetry action on the edge fields are:
\begin{align}
U(1)_c: &\begin{cases}
    \phi_\up \rightarrow \phi_\up+\alpha/4,\\
    \phi_\down\rightarrow \phi_\down-\alpha/4,
\end{cases} \nonumber\\
U(1)_{\rm sv}: &\begin{cases}
\phi_\up \rightarrow \phi_\up + \alpha/8,\\
\phi_\down \rightarrow \phi_\down + \alpha/8
\end{cases} \nonumber\\
\mathcal{T}: &\begin{cases}
\phi_\up\rightarrow -\phi_\down,\\
 \phi_\down \rightarrow -\phi_\up,
 \end{cases}.
\end{align}

The commutation relations: $[\phi_I(x),\d_y\phi_J(y)] = 2\pi iK^{-1}_{IJ}\delta(x-y)$ and charge assignments above, imply that the electrical charge and charge density, $Q,\rho(x)$, and spin/valley charge density $S^z,\rho_{\rm sv}$ operators are:
\begin{align}
Q &= \int dx \rho(x), &\rho(x) = \frac{1}{\pi}\d_x(\phi_\up-\phi_\down) \nonumber\\
S^z &= \frac12\int dx \rho_{\rm sv}(x),&\rho_{\rm sv}(x) = \frac{1}{\pi} \d_x(\phi_\up+\phi_\down)
\end{align}

The most general interaction parameter matrix consistent with these symmetries can be parameterized by a single dimensionless coupling constant $g$ as: $G_{IJ} = \delta_{IJ}+g\sigma^x_{IJ}$. 

The resulting scaling dimension for vertex operators $e^{ i(\phi_\up\pm\phi_\down)}$ are:
\begin{align}
\Delta_{\pm} = \frac{1}{8}\sqrt{\frac{1\mp g}{1\pm g}}
\end{align}

Similar to the $\Z_4$ order discussed above and in the main text, the superconducting boundary corresponds to imposing large term $-\lambda_S\int dx \cos 8(\phi_\up-\phi_\down)$ which pins $\phi_\up=\phi_\down +2\pi q/8$ to one of eight degenerate minima: $q=0,1,2,\dots 7$.
Similarly an IVC/SDW boundary corresponds to imposing a large term $-\lambda_M\int dx \cos 8(\phi_\up+\phi_\down)$ which pins $\phi_\up = -\phi_\down +2\pi s/8$ to one of eight degenerate minima $q=0,1,2,\dots 7$.
Similar to the $\Z_4$ case, the analog of the joint $e^2,m^2$-condensed boundary is complicated to describe in the Abelian bosonization framework.

Thus we see that magnetic or superconducting Cheshire qudits made from the helical $U(1)_8$ order will have $8$-fold GSD.

\subsubsection{Gapped boundaries: anyon condensation}
We proceed to show that any gapped boundary of the $\ueight$ order must condense the anyon $(4,4)$. Since condensing $(4,4)$ in $\ueight$ reduces it to the $\Z_4$ order, this will imply that the gapped boundaries of the $\ueight$ order are again in a one-to-one correspondence with those of the $\Z_4$ order (though, importantly, the resulting Cheshire qudits have different quantum dimensions and logical operator algebras).  
The bosons in the theory $\ueight$ are $(j,j), (j,-j),j=0,1\cdots, 7$ and $(4,0), (0,4)$. We divide Lagrangian anyon condensations in the $\ueight$ theory into two cases. 

\paragraph{Case I. $(4,0)$ or $(0,4)$ is condensed. } Say $(4,0)$ is condensed. Since $\theta_{(4,0),(j,0)}=e^{\frac{2\pi i}{8}4j}=(-1)^j$, the anyon $(1,0)$ is confined, while $(2,0)$ stays deconfined. Therefore the topological order is reduced to $U(1)_2\times U(1)_{-8}$. Bosons in this theory are $(0,0),(0,4),(2,\pm 2)$. Since all these four bosons are mutually bosonic, there is a unique Lagrangian condensation in  $U(1)_2\times U(1)_{-8}$ that condenses these four anyons. This means $(4,4)$ is condensed. 

\paragraph{Case II. $(4,0)$ and $(0,4)$ are not condensed.} Then the only other nontrivial bosons in the $\ueight$ theory are $(j,j),~(j,-j),~j=1,2,\cdots, 7$. It is clear that condensing any of these bosons implies condensation of $(4,4)$. 

Therefore, we have three gapped boundaries for the $\ueight$ order, corresponding to $e$, $m$ and $(e^2,m^2)$ condensation of the $\Z_4$ order. The explicit condensable algebras are listed below.
\begin{align}
    &\mathcal{A}^e=\sum_{j=0}^7 (j,j),\\
    &\mathcal{A}^m=\sum_{j=0}^7 (j,-j),\\
    &\mathcal{A}^{(e^2,m^2)}=(0,0)+(2,2)+(4,4)\nonumber\\
    &+(6,6)+(4,0)+(0,4)+(2,6)+(6,2).
\end{align}
 The defects between different gapped boundaries have identical topological properties to those of the $\Z_4$ order described above.

\end{document}